\documentclass[twocolumn,dvipsnames]{aastex63}
\usepackage[utf8]{inputenc}
\usepackage{comment}
\usepackage{booktabs,threeparttable}
\usepackage{changepage}
\usepackage{amsmath}
\usepackage{enumitem}
\usepackage{fontawesome5}
\usepackage{hyperref}

\makeatletter
\newcommand{\github}[1]{
  \href{#1}{\faGithubSquare}
}
\makeatother

\begin{document}
\title[\texttt{Mangrove}: Learning Galaxy Properties from Merger Trees]{\texttt{Mangrove}: Learning Galaxy Properties from Merger Trees}

\correspondingauthor{Christian Kragh Jespersen}
\email{ckragh@princeton.edu}

\author[0000-0002-8896-6496]{Christian Kragh Jespersen}
\affiliation{Department of Astrophysical Sciences, Princeton University, Princeton, NJ 08544, USA}

\author[0000-0002-6458-3423]{Miles Cranmer}
\affiliation{Department of Astrophysical Sciences, Princeton University, Princeton, NJ 08544, USA}

\author[0000-0002-8873-5065]{Peter Melchior}
\affiliation{Department of Astrophysical Sciences, Princeton University, Princeton, NJ 08544, USA}
\affiliation{Center for Statistics and Machine Learning, Princeton University, Princeton, NJ 08544, USA}

\author{Shirley Ho}
\affiliation{Department of Astrophysical Sciences, Princeton University, Princeton, NJ 08544, USA}
\affiliation{Center for Computational Astrophysics, Flatiron Institute, 162 5th Avenue, New York, NY 10010, USA}
\affiliation{Department of Physics, Carnegie Mellon University, Pittsburgh, PA 15217, USA}

\author[0000-0002-6748-6821]{Rachel S. Somerville}
\affiliation{Center for Computational Astrophysics, Flatiron Institute, 162 5th Avenue, New York, NY 10010, USA}

\author[0000-0003-4295-3793]{Austen Gabrielpillai}
\affiliation{Institute for Astrophysics and Computational Sciences, Catholic University of America, 620 Michigan Ave., DC 20064, USA}
\affiliation{Astrophysics Science Division, NASA/GSFC, 8800 Greenbelt Rd, Greenbelt, MD 20771, USA}
\affiliation{Center for Research and Exploration in Space Science and Technology, NASA/GSFC, 8800 Greenbelt Rd, Greenbelt, MD 20771, USA}

\begin{abstract}

Efficiently mapping baryonic properties onto dark matter is a major challenge in astrophysics. 
Although semi-analytic models (SAMs) and hydrodynamical simulations have made impressive advances in reproducing galaxy observables across cosmologically significant volumes, these methods still require significant computation times, representing a barrier to many applications. Graph Neural Networks (GNNs) have recently proven to be the natural choice for learning physical relations. Among the most inherently graph-like structures found in astrophysics are the dark matter merger trees that encode the evolution of dark matter halos. In this paper we introduce a new, graph-based emulator framework, \texttt{Mangrove}, and show that it emulates the galactic stellar mass, cold gas mass and metallicity, instantaneous and time-averaged star formation rate, and black hole mass ---as predicted by a SAM--- with root mean squared error up to two times lower than other methods across a $(75 Mpc/h)^3$ simulation box in 40 seconds, 4 orders of magnitude faster than the SAM. We show that \texttt{Mangrove} allows for quantification of the dependence of galaxy properties on merger history. We compare our results to the current state of the art in the field and show significant improvements for all target properties. \texttt{Mangrove} is publicly available.\footnote{ \url{https://github.com/astrockragh/Mangrove}} \\

\end{abstract}
 
\section{Introduction}

In the hierarchical paradigm of $\Lambda$CDM cosmology, dark matter is a crucial constituent of galaxy formation. While modeling the evolution of universes with only dark matter can be done both analytically \citep{Sheth2001halos} or through numerical N-body simulations \citep{Aarseth79_vearly_nbody, Efsthathiou85_nbody_historical, Maksimova21_abacus}, co-evolving dark matter and baryons still represents a major challenge, as no simple, direct mapping between the two exists \citep{Contreras15_nomap, deSanti22_DMBaryonMimick}. Instead we turn to simulations for modeling these complex interactions. Two widely accepted frameworks for doing so are semi-analytic models (SAMs) and hydrodynamic simulations, which in the last two decades have made it possible to populate cosmologically significant volumes with galaxies \citep{Somerville08_SAM, Vogelsberger14_Illustris, SomervilleDave2015, Naab17_nbody_galaxy_review, Vogelsberger20_GalaxyFormation_review}. However, the state-of-the-art hydrodynamical simulations take hundereds of millions of CPU hours to run \citep{Schaye15_EAGLE, Pillepich18_mainIllustrisTNG, Dave19_SIMBA, Paco21_CAMELS}. SAMs achieve much greater computational efficiency by combining dark matter \textit{merger trees} with a suite of physically motivated recipes for evolving the baryonic components of galaxies, but still require several hundreds of CPU hours to fill a $(75 Mpc/h)^3$ simulation box \citep{White91_hierarchical,Cole94_vEarlySAM,Somerville99_earlySAM, Benson12_galacticus, Lacey16_GALFORM, Lagos18_sharkSAM}.
Both methods reproduce many key observables over a broad redshift range, and semi-analytic and numerical hydrodynamic simulations make qualitatively similar predictions for many galaxy properties \citep{SomervilleDave2015}, as well as agreeing halo-by-halo on stellar and interstellar medium (ISM) properties \citep{Pandya:2020, ayromlou:2021,Gabrielpillai21_IllustrisSAM_comparison}. 

However, there are still significant differences between SAMs and hydrodynamic simulations, as well as tension between the behavior of individual SAMs and hydrodynamic simulations \citep{SomervilleDave2015}. This tension is seen both on a halo-by-halo basis and on a population-basis, depending on the parameter in question.

\citet{Kamdar16_SAM, Agarwal:2018, JoKim19_Illustris, Lovell22_hydroERT, deSanti22_DMBaryonMimick}(hereafter K16a, A18, JK19, L22, dS22) all attempt to map between dark matter and galactic baryonic properties using simple ML algorithms, like Extremely Randomized Trees, Random Forests, Multi-Layer Perceptrons or a combination of the above. These all attempt to map between dark matter halos and galactic properties using only features from the final halos at $z = 0$, or summary statistics believed to encode the merger history along with the features of the $z = 0$ halo. These methods all achieve reasonable success on select quantities such as galactic stellar mass and hot gas mass, but struggle to reconstruct quantities such as cold gas mass, star formation rate (SFR), and metallicity.
Even in cases where the regression methods were able to predict the \emph{median} values of a quantity with relatively low error (such as stellar mass), these techniques typically underestimated the \emph{dispersion} in the baryonic property at a given halo mass \citep{Agarwal:2018}. Galaxy properties such as stellar mass, ISM mass, stellar and ISM metallicity and SFR are known to lie in a relatively small sub-region of the high-dimensional parameter space; i.e., they populate a hyperplane. The ultimate goal of emulation methods is to reproduce \emph{the full hyperplane and its dispersion} for the full suite of baryonic properties of interest.

In this work, we present a new method for learning this highly non-trivial mapping, using the natural choice for learning on merger trees, a Graph Neural Network (GNN). GNNs have lately been demonstrated to work extremely well at modeling various problems in astrophysics \citep[e.g.,][]{Miles19_SymbolicGNNs, Miles20_symbolicregression,cranmerUnsupervisedResourceAllocation2021, cranmerHistogramPoolingOperators2021,Villanueva-Domingo21_gnnHalo,thielePredictingThermalSunyaevZel2022,lemosRediscoveringNewtonGravity2022}. This choice of model allows us to include the full merger history as recorded in the merger tree, since the merger tree can naturally be encoded as a graph.\\

As we will show, our model, \texttt{Mangrove}, outperforms all other models in the literature when predicting stellar mass, cold gas mass, black hole mass, cold gas metallicity, and SFR.\footnote{In addition to these 5, we also include SFR averaged over 100 Myr as a target variable.} This indicates that exploiting the inherent structure of the merger tree indeed is the stronger choice for mapping directly between dark matter and baryonic properties.
In this paper, we will furthermore demonstrate some valuable use cases that \texttt{Mangrove} allows for. First, we can probe the extent to which the exact merger history is important for predicting baryonic quantities, in a way that is infeasible with SAMs and hydrosimulations. Comparing results from using different parts of the merger trees allows us to quantify what aspects of baryonic properties are due to formation history, and which are due to the direct, time-independent dark matter-baryon connection, which has not been possible until now.

We can furthermore probe the importance of halo features along the merger tree in order to determine which parameters would be important in constructing an analytical theory that directly relates the merger tree to galactic properties, and where the most salient information lies. This is done by removing certain parameter sets and observing the change in model performance.

This paper is designed as follows. In \S\ref{sec:data} we present the simulations used to train \texttt{Mangrove}, as well as our data selection criteria. In \S\ref{sec:GNN} we briefly introduce GNNs and the loss function used for optimizing \texttt{Mangrove}. In \S\ref{sec:results} we present our results. In \S \ref{sec:high_z_targets} we present the results for predicting galactic stellar mass at different redshifts. \S\ref{sec:benchmark} compares our $z = 0$ results to the existing literature. \S\ref{sec:assembly_ablation} presents our exploration of the dependence of stellar mass on formation history, as well as an exploration of which dark matter features are most important for. In \S\ref{sec:discussionandfurther} we discuss our results and possible future work, and in \S\ref{sec:conclusion} we summarize our conclusions. An appendix with additional details is also provided.
\section{Simulations and Data}
\label{sec:data}

\subsection{Simulations and Merger Trees}
\label{sec:TNG}
We use the dark matter only version of the IllustrisTNG simulation, TNG-100-1-Dark. This simulation contains (1820)$^3$ particles within a box of $75 h^{-1}\mathrm{Mpc}$ on a side. This implies a dark matter particle mass of $6 \times 10^6$ h$^{-1} M_\odot$. The halo finding code {\sc rockstar} \citep{Behroozi13_rockstar} has been run on 99 snapshots from this simulation, and the {\sc consistenttrees} \citep{Behroozi13_consistenttrees} code is used to construct merger trees from these halo catalogues. See \citet[][hereafter G21]{Gabrielpillai21_IllustrisSAM_comparison} for more details on the halo finding and merger tree algorithms.

\begin{figure}[ht]
  \centering
  \includegraphics[trim=1.0cm 3cm 3cm 3cm, clip=true, width=\linewidth]{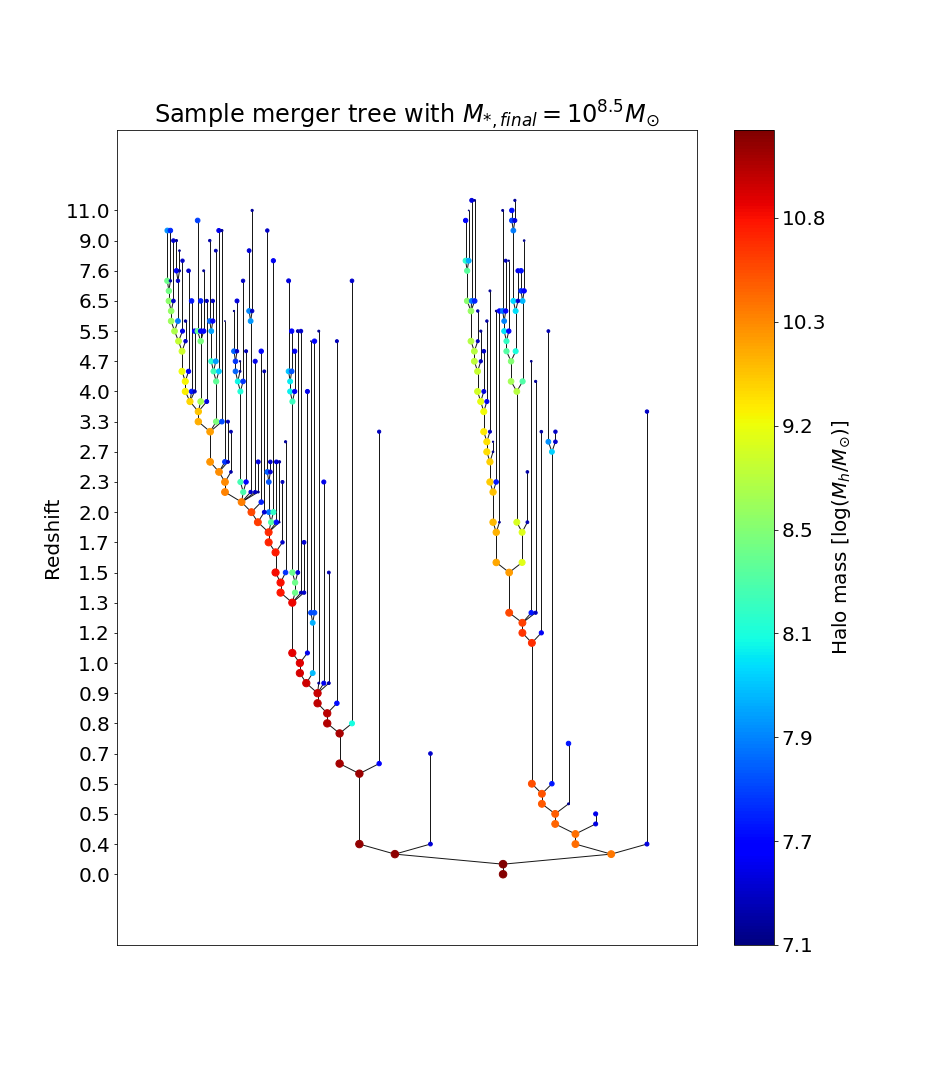}
  \caption{The merger tree of a $10^{8.5} M_{\odot}$ galaxy. Only nodes that are progenitors, merging or final (at $z = 0$), are shown. Both colour and dot size encode the mass of the halos. The merger tree encodes the full formation history and temporally evolving dark matter features of all halos that merge to form the final halo. Galaxies co-evolve with their dark matter halo, and the dark matter merger trees are therefore useful for determining galaxy formation.}
  \label{fig:full_tree}
\end{figure}

\subsection{Santa Cruz Semi-Analytic Model}
\label{sec:SAM}
We then run the well-established Santa Cruz Semi-Analytic Model \citep{Somerville99_earlySAM, Somerville08_SAM,somerville:2015} on the merger trees described above. The current version of the SC-SAM is documented in G21. 

Fundamentally, the SAM uses a set of coupled ODEs to track the flow of matter, gas, and metals between different reservoirs (the diffuse intergalactic medium, hot gas halo, cold gas in the ISM, stars, etc.).
The predictions of the SC-SAMs have been extensively compared with observations from $z\sim 0-10 $, and the SC-SAMs predictions are in good agreement with observables such as stellar mass functions, SFR distributions, and cold gas content \citep{Somerville08_SAM,somerville:2015,yung:2019a,yung:2019b, somerville:2021}. Moreover, they produce reasonably similar predictions to hydrodynamic simulations for the stellar mass content and cold ISM content of galaxies across a broad range of halo mass and cosmic time. See \citet{SomervilleDave2015} for a review comparing many SAMs and hydrodynamical simulations, \citet{Pandya:2020} for a comparison of the Santa Cruz SAM with the FIRE simulations, and G21 for a comparison of the SC-SAMs with the IllustrisTNG simulations.

The SAM reproduces the galaxy ``assembly bias'' (dependence of clustering on properties other than halo mass) seen in IllustrisTNG, which is by definition not reproduced by regular Halo Occupation Distribution (HOD) models \citep{Hadzhiyska:2021}. 

\subsection{Data Selection}

To ensure that the galaxies in the dataset have reliable features and targets, we employ a set of selection criteria. First, only merger trees where the final halo has a mass of $10^{10} M_{\odot}$ or above are included. This choice is made as the mass of the final halo indicates both the reliability with which the dark matter properties can be measured as well as the reliability of the derived SAM baryonic properties. Secondly, only central galaxies are included, since centrals and satellites are believed to have different relationships with their host halos \citep{Hearin16_decoratedHOD}. An inclusion of satellites in future work would be of great interest.

Since in any given merger tree there can be upwards of millions of nodes, some reductions are made. As we are mainly interested in probing the merger history, we preserve nodes/halos that are either: \begin{itemize}
  \item A progenitor node, i.e., the first time a halo was detected in the simulation
  \item Pre-merger nodes, i.e., halos from the snapshot before they merge
  \item Post-merger nodes, i.e., halos that are the direct result of a merger
  \item The final node, i.e., the final halo
\end{itemize} 

This reduces the number of nodes by a factor of $\sim$10-50, depending on the merger tree in question. A sample merger tree resulting from this selection can be seen in Figure \ref{fig:full_tree}. Our selection produces a strong inductive bias, since smooth accretion modes are not included. We also limit the total number of nodes to be $< 2\cdot10^4$, which results in the exclusion of 107 merger trees. Since we regress logarithmic targets, only galaxies with non-zero target quantities are included, excluding 470 trees. In total, the $z=0$ dataset consists of 108,338 merger trees. In the future, this problem should be addressed more eloquently, either by combining \texttt{Mangrove} with a classifier predicting if a given galaxy has a zero-value target or not. Another transformation such as the arcsinh transformation could also be considered.

In the basic dataset, we include all dark matter features that are not IDs, x,y,z positions, or x,y,z velocities, even features not explicitly used by the SAM. See \S \ref{asec:params} in the appendix for a list of halo parameters. The description of our training, validation, and test set selection can also be found in \S \ref{asec:train_val_test} of the appendix.

SAMs output a large range of baryonic galactic properties, but for exploring the possibility of emulating them with a GNN, we pick a few quantities of interest. The main target of interest is \textit{stellar mass} ($\rm log(M_*/M_{\odot})$, hereafter $M_*$). This is a central quantity for both creating mock catalogues and for simulators to successfully reproduce, and is therefore also the main focus of this project.

To explore the possibility of emulating other baryonic properties that are physically significant or closely related to observables, we also include a range of other targets.

\begin{itemize}
  
  \item Cold gas (ISM) mass ($\rm log(M_{cold}/M_{\odot})$, hereafter $M_{cold}$). The cold gas is the fuel for star formation, and can also be probed observationally through sub-mm emission lines such as CO and through the dust continuum. This is also explored by A18 and L22.
  \item Black hole mass ($\rm log(M_{BH}/M_{\odot})$, hereafter $M_{BH}$). The supermassive black hole at the center of a galaxy influences the entire galaxy way beyond its gravitational influence \citep{Kormendy13_blackhole_galaxy}. This is also explored by JK19 and L22.
  
  \item Cold gas (ISM) metallicity ($\rm log(M_{Z_{gas}}/M_{cold}))$, hereafter $Z_{gas}$). The cold gas metallicity is observable through the strength of different metal lines and is used as a tracer of cold gas clouds. This quantity is also explored by A18 and L22.
  
  \item Instantaneous Star Formation Rate ($\rm log(SFR/M_{\odot}/yr)$, hereafter $SFR$). This is also explored by A18, JK19, L22 and dS22.
  
  \item Star Formation Rate averaged over 100 Myr ($\rm log(SFR_{100}/M_{\odot}/yr)$, hereafter $SFR_{100}$). Both SFR properties can be probed through observations of UV or FIR light, and emission lines such as H$\alpha$, and correlate strongly with color. We attempt to regress both the current and 100 Myr averaged SFR, since these in conjunction should provide crucial information about the recent merger history of a galaxy \citep{Caplar19_SFMS, Iyer20_SF_variability}. 
\end{itemize}
\vspace{1cm}

\section{Graph Neural Networks}
\label{sec:GNN}

\begin{figure*}
  \begin{adjustwidth}{-.2cm}{-.4cm}
  \centering
  \includegraphics[trim=0cm 4.2cm 0cm 7cm, clip=true, width=0.9\linewidth]{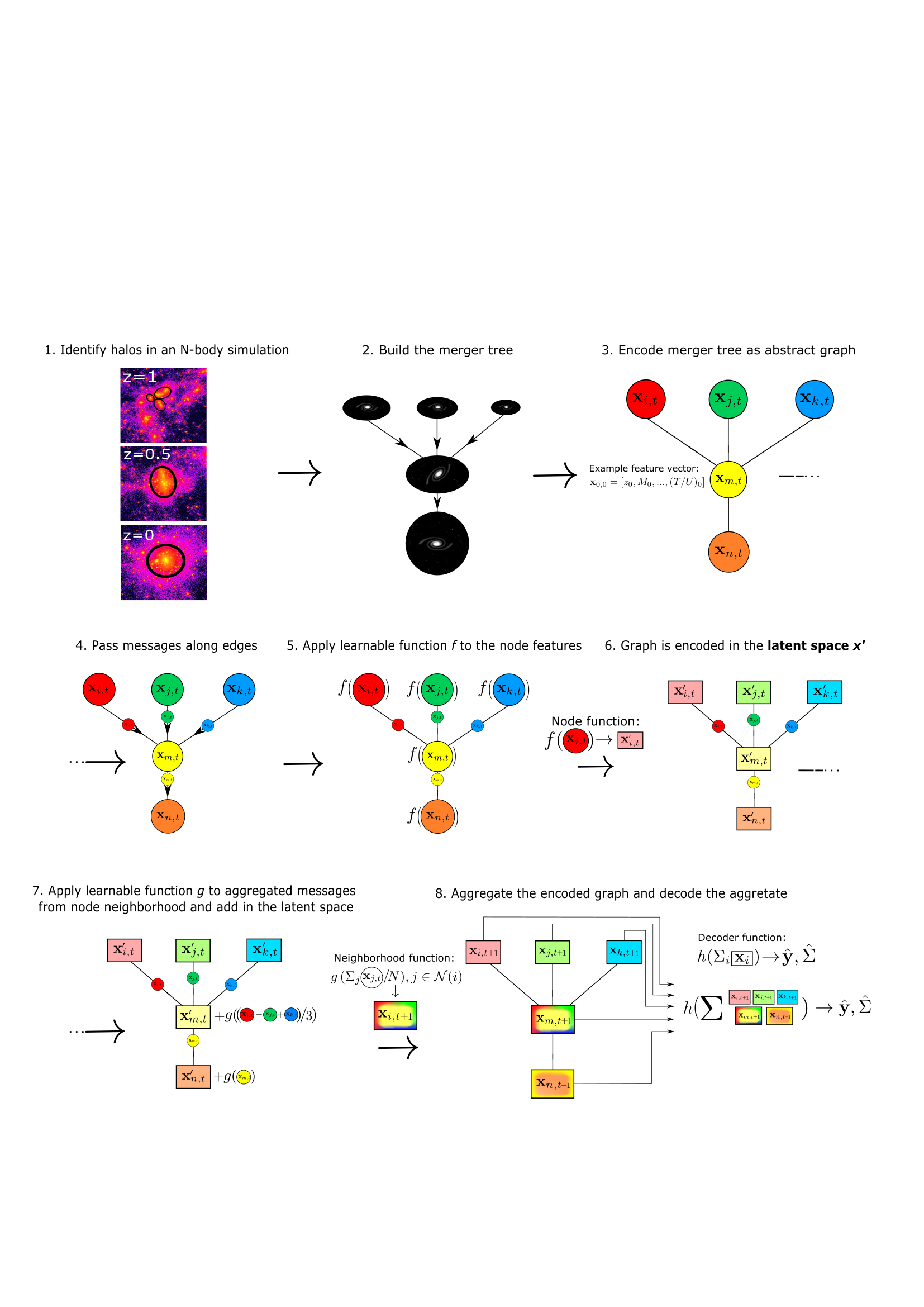}
  \end{adjustwidth}
  \caption{An illustration of the full workflow of \texttt{Mangrove}. Merger trees are encoded as graphs, which are then passed trough \texttt{Mangrove}. Messages are passed forward in time only, since merger trees are directed in time. Node states are updated by applying a learnable function \textit{f} to the current node states, applying a learnable function \textit{g} to the mean of the node states of the neighboring nodes and adding these two as described in \S \ref{sec:GNN}. Transferring into the latent space is marked by shaded colors and a change of shape from circles to rectangles. Adding neighborhood information is marked by mixing of colors. The model makes message-passing steps, after which the graph nodes are summed over, and this sum is then decoded by another learnable function, \textit{h}, which gives the predictions and the Gaussian covariance matrix. All learnable functions are \textbf{Multi-Layer Perceptrons}.}
  \label{fig:GNN}
\end{figure*}

The most successful models are the ones which embed well-motivated inductive biases into the model that one wishes to fit to some data. In machine learning, using convolutional neural networks (which preserve locality) for images is a prime example of this general principle. Since halo and galaxy evolution are naturally encoded in merger trees, which are graphs, a Graph Neural Network (GNN) is an intuitive choice for mapping galactic baryonic physics onto dark matter. GNNs, as other ML methods, are built from a sequence of modules, called \textbf{layers}. These layers are then stacked as a sequential series of message-passing or graph convolutional layers \citep{KipfWelling17_GraphConv, Battaglia18_GNN_overview} which pass information from the \textit{nodes} along the \textit{edges} of the graph, followed by a differentiable pooling function and a decoder function, which is usually a Multi-Layer Perceptron (MLP) \citep{Rumelhart86_MLP}. The pooling function is applied in order to standardize the outputs for graphs varying number of nodes. The overall flow of \texttt{Mangrove} is visualized in Figure \ref{fig:GNN}.

A description of the full structure of \texttt{Mangrove}, known as the \textbf{architecture} of the model, can be found in \S \ref{sec:architecture} in the appendix.

\subsection{Core Concepts}
GNNs are a species of neural network which operate on graph-structured data \citep{Scarselli2008_GNN, Bronstein17_GeometricDL,Battaglia18_GNN_overview}. For our purpose, the graphs, $G$, on which GNNs operate are defined as 2-tuples, $G=(V, E)$,\footnote{We adhere closely to the notation used in \cite{Battaglia18_GNN_overview, Miles19_SymbolicGNNs} for formal definitions.} where $V=\left\{\mathbf{v}_{i}\right\}_{i=1: N^{v}}$, and $N^{v}$ is the total number of nodes, is a set of node attribute vectors of dimensionality $D^{v}$. The edges can be encoded by $E=\left\{\left(\mathbf{e}_{k}, r_{k}, s_{k}\right)\right\}$, a set of edge attribute vectors of dimensionality $D^{e}$, and with indices $r_{k}, s_{k} \in\left\{1: N^{v}\right\}$ of the ``receiving" and ``sending" nodes connected by the $k$-th edge.

In this work, only node attributes and edge indices are used, although edge features could be created. Note that our graphs are \textbf{directed}, since merger trees are inherently directed in time. A directed graph means that information can only be passed one way on a given edge, which for our purpose follows the flow of time since propagating information backwards in time would break causality.

The \textbf{neighborhood} of node $i$ consists of all nodes that are connected to node $i$ by an edge. Note that for a directed graph, this only includes the set of nodes for which $r_k=i$. Some prefer to instead define two separate notions of neighborhoods for directed graphs, an \textbf{incoming} neighborhood and an \textbf{outgoing} neighborhood. Our definition would be the same as the incoming neighborhood. We denote the neighborhood of node $i$ by $\mathcal{N}(i)$.

\subsection{GraphSAGE}
\label{sec:conv_layer}

As described above, the message-passing or graph convolutional layer makes up the cornerstone of GNNs. Information from the neighborhood is passed along the edges leading to a given node, in order to learn not just from the node feature vector but from its neighbors.

There has been a variety of different proposed message-passing and graph convolutional layers. In this project we use the \verb|PyTorch Geometric| \citep{Fey19_PyTorchGeometric}, implementation of the GraphSAGE convolutional layer from \cite{Hamilton17_GraphSAGE}. With each application of this layer, each node updates its state from the input state $\mathbf{v}_i$ to a \textit{hidden state} $\mathbf{v}^{\prime}_i$, through:
\begin{equation}
  \mathbf{v}^{\prime}_i = \mathbf{W}_1 \mathbf{v}_i + \mathbf{W}_2 \cdot
    \mathrm{mean}_{j \in \mathcal{N}(i)} \mathbf{v}_j 
\end{equation}

where $\mathbf{W}_1$ and $\mathbf{W}_2$ are learnable weight matrices. Thus, $\mathbf{W}_1$ operates on information from the node itself, and $\mathbf{W}_2$ on the mean\footnote{Although the mean is used here, any other differentiable aggregation function is usable.} of the node states of the neighborhood nodes. 

In order to gain some intuitive understanding of what this layer actually does, it is useful to think about these layers in terms of function. We can introduce two learnable functions, $f$, the \textbf{node function} and $g$, the \textbf{neighborhood function}, which are both constrained to be linear. Then a full application of GraphSAGE can be written as:

\begin{equation}
  \mathbf{v}^{\prime}_i = f(\mathbf{v}_i) + g(\mathrm{mean}_{j \in \mathcal{N}(i)} \mathbf{v}_j )
\end{equation}

The optimization task can then be framed as learning the functions $f$ and $g$, expressed through matrices $\mathbf{W}_1$ and $\mathbf{W}_2$. See Figure \ref{fig:GNN} for a schematic of the flow of these learnable functions.

Since all functions in this layer are linear, a non-linear \textit{activation function}, usually written as $\sigma$,\footnote{Not to be confused with the scatter or standard deviation.} is applied between each layer, allowing expression of nonlinear functions. Thus the node state is updated to:
\begin{equation}
  \mathbf{v}^{\prime}_{i+1} = \sigma(\mathbf{v}^{\prime}_i)
\end{equation}

In this work we use the ReLU activation function between GraphSAGE layers \citep{Agarap18_ReLU}. 

\subsection{Loss Function}
\label{sec:loss_func}

The loss function $\mathcal{L}$ is central to the optimization and performance of any GNN, as the parameter set $\theta$ which makes up the GNN is optimized to satisfy $min(\mathcal{L}_{\theta}(\{G\}_{train}))$. In this work we employ a generalized Gaussian Negative Log-Likelihood (NLL).

For a single input, the general Gaussian NLL is defined as:

\begin{equation}
  \mathcal{L}(\mathbf{y}, \mathbf{\hat{y}}, \hat{\Sigma}) \equiv
  \frac{ln(|\hat{\Sigma}|)}{2} +\frac{1}{2}\left(\mathbf{y}-\mathbf{\hat{y}}\right)^{T} \hat{\Sigma}^{-1}\left(\mathbf{y}-\mathbf{\hat{y}}\right)
\end{equation}
where, $\mathbf{y}$ is the true target vector, $\mathbf{\hat{y}}$ is the network prediction vector, $\hat{\Sigma}$ is the predicted covariance matrix, and $|\hat{\Sigma}|$ denotes its determinant. These are then easily extended to their batch form, by simply summing over all inputs in a batch.

In this paper, the quoted results are obtained via a purely diagonal covariance matrix, although in some cases, we obtained better results with the full covariance matrix. The diagonal covariance matrix is preferred for its simplicity, and still renders Gaussian uncertainties, which are highly useful as discussed in \S \ref{sec:discussionandfurther}. The metrics (see \S \ref{sec:metrics}) between the two cases differ by no more than a few \%.
A further discussion of the uncertainties can be found in Appendix \S \ref{asec:uncertainties}.

\section{Results}
\label{sec:results}

In this section, we first introduce the metrics used to characterize the performance of \texttt{Mangrove}. We then present results for our $M_*$ predictions and compare them to other methods and the possibility of generalization across different redshifts. We then explore the predictions of the other target parameters.\\
We compare our results to results from four other frameworks.
\begin{itemize}
\itemsep0em
    \item Our $M_*$ prediction will be compared to the more widely used method for connecting halo masses and galactic stellar masses, Abundance Matching \citep{ValeOstriker04_AbundanceMatching}.
    \item Where possible, we compare to other papers in the literature which have attempted to regress the same target parameters. However, these are not performed on the same dataset as ours.
    \item In order to mitigate this issue, we train a MLP on the $z=0$ halos (final halos) of our dataset which should be comparable to the methods in the literature.
    \item As a way of providing an estimate of the best possible performance on the test set, we run the SC-SAM with different random seeds, and calculate the comparison metrics between them. This should be an estimate of the lower information limit. If our predictions do significantly better than this across the entire simulation, it would be cause for some concern. This is not the same as the numerical uncertainty, but arises due to the fact that while the model can learn summary statistics for the probability distributions used in the SAM, it is a fully deterministic model and thus could not emulate the random draws. Going forward this is denominated as the \textbf{SAM probabilistic limit}.
\end{itemize}
\subsection{Metrics}
\label{sec:metrics}
The most commonly used metric to determine accuracy in astronomy is the scatter\footnote{Also commonly referred to as the standard deviation or root-mean-squared-error.}, defined as:
\begin{equation}
  \sigma \equiv \sqrt{\frac{1}{N_{test}} \sum^{N_{test}} (\Delta y - \overline{\Delta y}) ^2}
\end{equation}

where $\Delta y \equiv y - \hat{y}$ is the residual of a single prediction and $\overline{\Delta y}$ is the mean of the residuals.\\
This metric has two significant caveats. Firstly, it does not measure any systematic offset in the residuals. Therefore, as an important addition, we introduce the bias as an auxiliary metric, defined as the mean of the residuals, i.e.:
\begin{equation}
  \text{bias} \equiv \frac{1}{N_{test}} {\sum^{N_{test}} \Delta y}
\end{equation}

The bias effectively measures any systematic offset. The best possible predictions would have low scatter and no bias. \\
Second, because the scatter is susceptible to outliers, we include two secondary metrics that are more stable and not directly optimized for in our loss function.

\begin{itemize}[noitemsep]

  \item Pearson correlation coefficient ($\rho$), i.e., the linear correlation between the target and model prediction: 
  \begin{equation}
    \rho \equiv \frac{\operatorname{cov}\left(y, \hat{y} \right)}{\sigma_{y} \sigma_{\hat{y}}}
  \end{equation}
  
  \item Coefficient of determination ($R^2$):
    \begin{equation}
    R^2 \equiv 1-\frac{\sum (\Delta y)^2}{\sum (y-\overline{y})^2} 
  \end{equation}
\end{itemize}

For both of these metrics, a perfect set of predictions would correspond to $\rho_{Pearson}=R^2=1$ 

It is important to note that the ability to predict any specific target generally improved when \texttt{Mangrove} was trained to predict all target variables. 
We therefore distinguish between models trained for all targets and only a single target when presenting results.

\subsection{Stellar Mass Results}

\begin{figure*}
  \begin{adjustwidth}{-1.2cm}{-.2cm}
  \centering
  \includegraphics[width=0.95\linewidth]{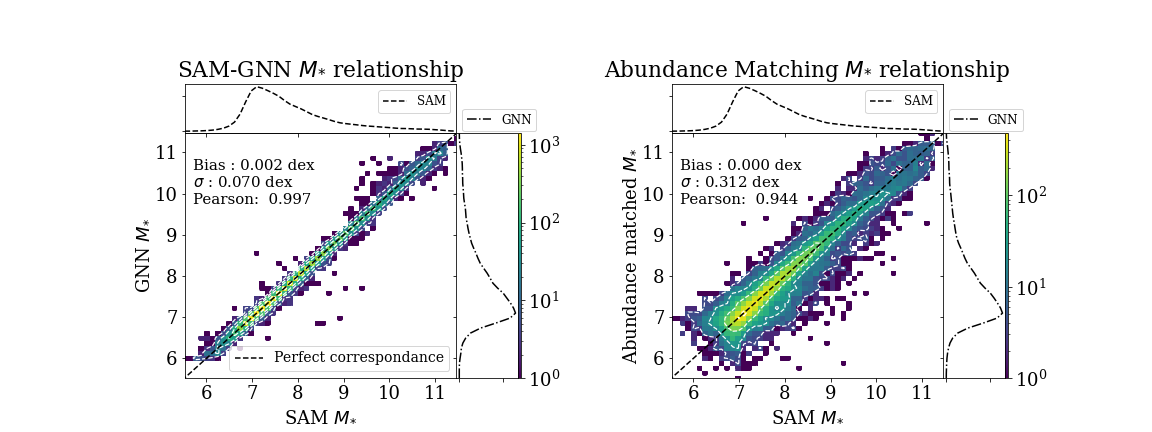}
  \end{adjustwidth}
  \caption{Histogram of the SAM $M_*$ versus predicted $M_*$ with logarithmically colored bin heights. The left panel shows the target-prediction relation for our GNN, \texttt{Mangrove}, and the right panel shows target-prediction relation of the common abundance matching approach. 
  Kernel Density Estimates of the SAM and \texttt{Mangrove} - predicted distributions are shown on the relevant axes. By leveraging the formation history of the galaxy via the merger tree, we obtain precise and accurate predictions of $M_*$. The prediction scatter is improved by almost a factor of two compared to the state of the art, single-halo method and by more than a factor of 4 compared to the widely used abundance matching method (see Table \ref{tab:comparison_results}). \texttt{Mangrove}'s $M_*$ scatter is comparable to, but still well above, the probabilistic limit of the SAM, which is at 0.043 dex.} 
  \label{fig:mstar_performance}
\end{figure*}

As the central quantity of interest, the stellar mass received the most attention in this paper. The test set results were a scatter of \textbf{0.070 dex}, with \textbf{0.002 dex} bias. This is shown in Figure \ref{fig:mstar_performance}, along with a comparison to the usual halo mass abundance matching approach.\footnote{Other metrics can be found in Table \ref{tab:comparison_results}}
Abundance matching \citep{ValeOstriker04_AbundanceMatching}, simply rank-orders all galaxies and halos by mass and assumes a monotonic matching relation exists between the two. We include this comparison as a baseline due to its simplicity and widespread use.

Figure \ref{fig:mstar_performance} shows the relation between target value and predicted value, along with distributions on the respective axes. The Figure shows the (target, prediction) - relation as a 2D histogram with logarithmic bin heights. If this relation follows the diagonal, that would indicate perfect predictions. The tighter the relation follows the diagonal, the better. 

A few comparisons are beneficial to keep in mind:
\begin{itemize}
  \item Training \texttt{Mangrove} to predict only $M_*$ yields a scatter of 0.078 dex, 11\% worse than the performance when training \texttt{Mangrove} to predict all quantities simultaneously.
  \item The performance of the model worsens to a scatter of 0.132 dex when using only the parameters of the final halo, indicating a strong dependence on assembly history.
  \item The scatter of \texttt{Mangrove}'s $M_*$ predictions is comparable to the SAM probabilistic limit as defined above, which has a scatter of 0.043 dex (see Table \ref{tab:comparison_results}).
\end{itemize}

Further investigation of the impact of different features and the merger history can be found in \S \ref{sec:assembly_ablation}.

\subsubsection{Stellar mass at other redshifts}
\label{sec:high_z_targets}

\begin{figure}
  \begin{adjustwidth}{-0.8cm}{-0.1cm}
  \centering
  \includegraphics[width=.93\linewidth]{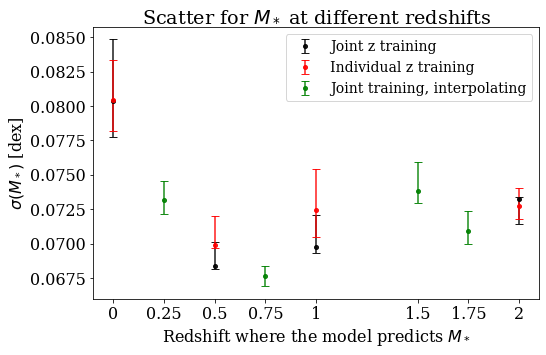}
  \end{adjustwidth}
  \caption{Median, 16th and 84th percentile of 10 models trained to predict only $M_*$ at a series of different redshifts in three different ways. Red is when testing at the same single redshift where \texttt{Mangrove} was trained, black is from training and testing jointly at $z \in {0,0.5,1,2}$ and green is from training jointly at $z \in {0,0.5,1,2}$ and testing at $z \in {0.25,0.75,1.5,1.75}$. \texttt{Mangrove} predicts $M_*$ at previously unseen redshifts with similar or lower scatter than the redshifts at which \texttt{Mangrove} was trained.}
  \label{fig:mstar_z_generalization}
\end{figure}

\begin{figure}
  \begin{adjustwidth}{-0.8cm}{-0.1cm}
  \centering
  \includegraphics[width=0.9\linewidth]{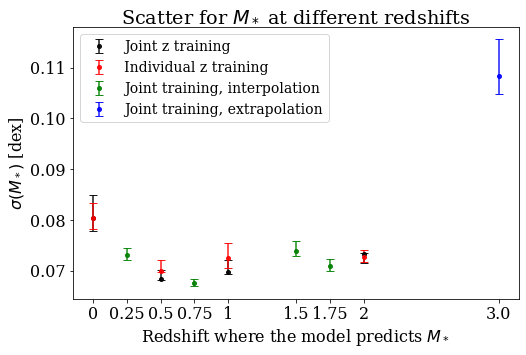}
  \end{adjustwidth}
  \caption{Same as Figure \ref{fig:mstar_z_generalization}, but including a point from extrapolating to $z=3$ using the models trained at $z \in {0,0.5,1,2}$. The predictions have much higher scatters.}
  \label{fig:mstar_z_extrapolation}
\end{figure}

A central question for many ML models is to what extent its predictions will \textit{generalize}. For astrophysical purposes, generalization across redshifts is crucial. We investigated whether \texttt{Mangrove} would perform well at $z>0$ by doing several experiments.
\begin{enumerate}
  \item Test models trained at a single redshift at redshifts where they were not trained. This in general leads to imprecise and highly biased results.\footnote{bias and $\sigma>0.1$}
  \item Train and test models at individual $z \geq 0$. \texttt{Mangrove} can achieve an accuracy below 0.08 dex at all redshifts as can be seen in Figure \ref{fig:mstar_z_generalization}.
  \item Train a general model by pooling training sets at $z \in {0,0.5,1,2}$, and testing at $z \in {0,0.5,1,2}$. Compared to training and testing at individual redshifts, we obtain similar results at all redshifts.
  \item \textit{Most surprisingly}, by pooling training sets at $z \in {0,0.5,1,2}$, and testing at $z \in {0.25,0.75,1.5,1.75}$ where \texttt{Mangrove} was \textit{not} trained, we obtain comparable bias and scatter to where \texttt{Mangrove} was trained. 
  \item Pooling training sets at $z \in {0,0.5,1,2}$, and testing at $z = 3$ where \texttt{Mangrove} was not trained, we get a relatively inaccurate result as seen in Figure \ref{fig:mstar_z_extrapolation}.
\end{enumerate}

All experiments in this section were done with models trained to predict only $M_*$.

\texttt{Mangrove} performs well on all $z>0$, even learning a smooth transformation for general redshifts if in the interpolative regime. However, it worsens significantly when extrapolating.

\subsubsection{$M_{cold}$, $M_{BH}$, $Z_{gas}$, $SFR$ and $SFR_{100}$ Results}

\begin{figure*}
  \begin{adjustwidth}{-0.1cm}{-0.2cm}
  \centering
  \includegraphics[width=1\linewidth]{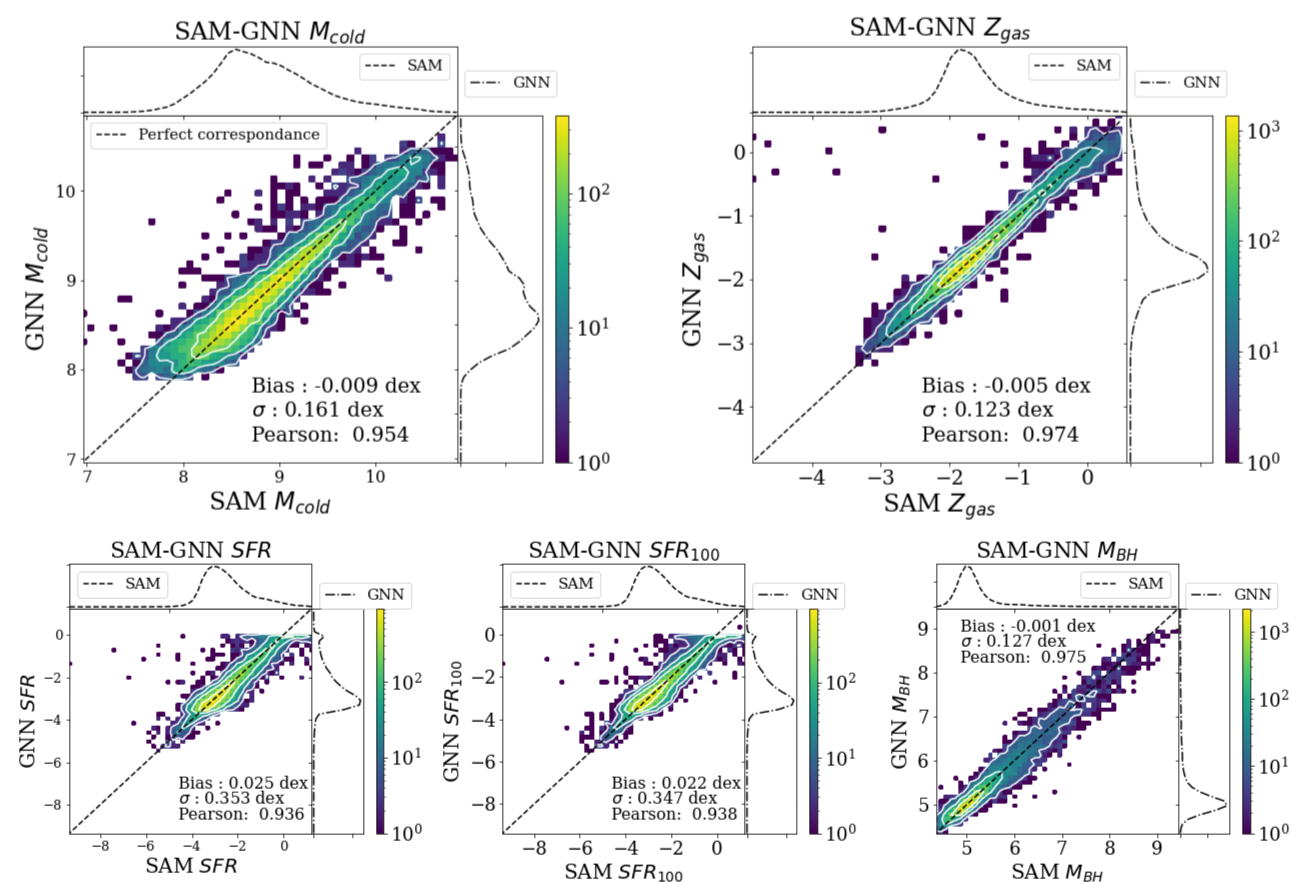}
  \end{adjustwidth}
  \caption{Histogram of SAM targets versus the predicted targets of our GNN, \texttt{Mangrove}, with logarithmically colored bin heights. Recovering galactic $M_{cold}$, $Z_{gas}$, $SFR$, $SFR_{100}$ and $M_{BH}$ renders mostly tight and unbiased relationships across the entire ranges of these galactic features, with the notable exception of targets where $SFR$ or $SFR_{100} > 0.1$. This flaw is spurious and should be possible to mitigate. All of these results outperform results from the literature for all precision metrics (see Table \ref{tab:comparison_results}). The fact that such mappings exist imply that the full hyperplane of galactic properties may be emulated with \texttt{Mangrove}.}
  \label{fig:others_performance}
\end{figure*}

In general, we observe a weaker dependence on merger history for all other target quantities, but the predictions from using the full merger tree are always significantly better than just using the final halo, as can be seen in Table \ref{tab:comparison_results}.\footnote{This can also easily be seen if comparing Figures \ref{fig:mstar_performance} and \ref{fig:others_performance} to Figure \ref{fig:final_halo} in the appendix.} The regression results for $M_{cold}$, $M_{BH}$, $Z_{gas}$, $SFR$ and $SFR_{100}$ for a model trained to predict all targets at the same time are visualized in Figure \ref{fig:others_performance}. As can be seen, \texttt{Mangrove} generally performs very well but struggles in regions of $SFR$ or $SFR_{100} > 0.1$. This is due to the two diverging branches in SFR that occur around this value (see Figure \ref{fig:Mhalo-targets} in the appendix).

Notably, the $M_{BH}$ predictions are slightly better than the SAM probabilistic limit. Since the predictions are just below this limit, we can interpret the results as having accurately captured a generalizing rule for mapping between the dark matter merger tree and $M_{BH}$, with little cause for concern.

See \S \ref{asec:more_plots} in the appendix for an in-depth analysis of both the relations between halo mass and the target parameters, as well as the interdependence of the residuals of the target parameters. 

\subsection{Comparison to Benchmark}
\label{sec:benchmark}
\noindent
\begin{table*}
\begin{adjustwidth}{-0.5cm}{}
\begin{threeparttable}
\begin{tabular}{l l cc cc cc cc } 
\toprule
Target &  Paper/Method & $\sigma$ [dex] & Bias [dex] & $\rho_{Pearson}$ & $R^2$ - score \\ 
\midrule
 $M_*$ & \texttt{Mangrove} & \textbf{0.070} & 0.002 & \textbf{0.997} & \textbf{0.994}  \\
 & Final halo only & 0.132 & 0.003 & 0.990 & 0.980 \\
 & A18 & 0.189 & 0.004 & 0.953 & 0.909  \\
 & JK19 & 0.162 & 0.027 & 0.991 & 0.982 \\
 & dS22 & 0.132 & - & 0.977 & 0.954  \\
 & L22 & - & - & 0.924 & -  \\
 & Abundance Matching & 0.312 & \textbf{0.000} & 0.944 & 0.889  \\
 & SAM probabilistic limit & 0.043 & 0.000 & 0.999 & 0.997 \\
\midrule
$M_{cold}$ & \texttt{Mangrove} & \textbf{0.161} & -0.009 & \textbf{0.954} & \textbf{0.909} \\
 & Final halo only & 0.182 & \textbf{0.001} & 0.941 & 0.885 \\
& L22 & - & - & 0.799 & -  \\
& SAM probabilistic limit & 0.096 & 0.000 & 0.983 & 0.967 \\
\midrule
$M_{BH}$ & \texttt{Mangrove} & \textbf{0.127} & \textbf{-0.001} & \textbf{0.975} & \textbf{0.950} \\
 & Final halo only & 0.175 & -0.013 & 0.951 & 0.904  \\
 & JK19 & 0.272 & 0.044 & 0.965 & 0.927 \\
& L22 & - & - & 0.881 & - \\
& SAM probabilistic limit & 0.129 & 0.000 & 0.967 & 0.934 \\
\midrule
$Z_{gas}$ & \texttt{Mangrove} & \textbf{0.123} & \textbf{-0.005} & \textbf{0.974} & \textbf{0.948} \\
 & Final halo only & 0.151 & -0.007 & 0.960 & 0.922 \\
 & A18 & 0.160 & 0.004 & 0.860 & 0.739 \\
& SAM probabilistic limit & 0.074 & 0.000 & 0.990 & 0.981 \\
\midrule
$SFR$& \texttt{Mangrove} & \textbf{0.353} & -0.025 & \textbf{0.936} & \textbf{0.876} \\
& Final halo only & 0.392 & \textbf{0.002} & 0.921 & 0.847 \\
 & A18 & 0.433 & -0.014 & 0.745 & 0.555  \\
 & JK19 & 0.93 & -0.024 & 0.902 & 0.760 \\ 
& dS22 & 0.850 & - & 0.652 & 0.094  \\
& L22 & - & - & 0.804 & -  \\
& SAM probabilistic limit & 0.230 & 0.000 & 0.973 & 0.947 \\
\midrule
$SFR_{100}$& \texttt{Mangrove} & \textbf{0.347} & 0.022 & \textbf{0.938} & \textbf{0.879} \\
& Final halo only & 0.388 & \textbf{0.003} & 0.922 & 0.849 \\
 & SAM probabilistic limit & 0.200 & 0.000 & 0.980 & 0.960 \\
\bottomrule
\end{tabular}
\end{threeparttable}
\end{adjustwidth}
\caption{Metrics for the methods discussed in this paper. SAM probabilistic limit denominates the metrics obtained from comparing separate realizations of the SC-SAM. \texttt{Mangrove} denominates the results of \texttt{Mangrove} using the full merger history and all halo parameters. Final halo only denominates the results of \texttt{Mangrove} using all halo parameters for the $z=0$ halo, i.e., the final halo. We bold the best emulator performance for each metric for each target variable. Exact values from JK19, dS22 and L22 were obtained through correspondence with the authors. Only for bias are our results comparable to results from the literature, although the bias is usually small enough to be dominated by noise. Note that although the $SFR$/$SFR_{100}$ scatter is quite high, so is the SAM probabilistic limit. For $M_{BH}$, \texttt{Mangrove} notably performs slightly better than the SAM limit.}
\label{tab:comparison_results}
\end{table*} 

To evaluate our performance against the current state-of-the-art in mapping baryonic properties directly onto dark matter, we provide Table \ref{tab:comparison_results}, containing accuracy metrics\footnote{See beginning of \S \ref{sec:results} for an overview of the used metrics} for this work and others in the literature at $z=0$. 

Although we here compare directly to the literature results, it should be noted that the dS22 report values for \textit{specific, instantaneous} SFR, sSFR\footnote{Defined as $SFR/M_*$.}, which means that their results for just SFR have significantly higher scatter, since this prediction is dominated by the stellar mass prediction.

Since papers from the literature do not report all metrics used here, we provide an estimate for a final-halo-only result, which effectively follows the basic Neural Network approach of dS22. The final halo only accuracy should also be comparable to the approaches using tree-based algorithms, as shown by dS22, so the results of A18, JK19 and L22 should also be comparable.

As described in \S \ref{sec:results}, we also provide the probabilistic limit of the SAM.

Comparing our results to results from the literature, we see that \texttt{Mangrove} outperforms other models across all categories, as well as showing remarkable improvements for using the full structure compared to only using the properties of the final halo. Especially $M_*$ and $M_{BH}$ show very significant improvements comparing our merger tree approach with the final halo only approach.

\section{Dependence on assembly history and feature ablation}
\label{sec:assembly_ablation}

A key method of investigating ML models is to simply remove certain parts of either the model or the input data, in order to probe the importance of said part. This process is known as \textit{ablation}. In this section we seek to quantify the impact on our model accuracy by removing parts of the input data. We here investigate two separate cases, in both cases with models trained to predict only $M_*$ at $z=0$.
\begin{itemize}
  \item The dependence of our model accuracy on the fraction of the merger history included in the input merger trees. 
  \item The dependence of our model accuracy on a different sets of input features when given the full merger tree as an input. 
\end{itemize}

These tests investigate what aspects of current galactic stellar masses are due to history, and which are inevitable due to the fundamental physical connection between different aspects of a galaxy and its host halo, and which of these aspects are the most important.

\subsection{Dependence on Merger History}
\label{sec:assembly_bias}

\begin{figure}[ht]
  \centering
  \includegraphics[trim=1.1cm 3cm 3cm 3.cm, clip=true,width=1\linewidth]{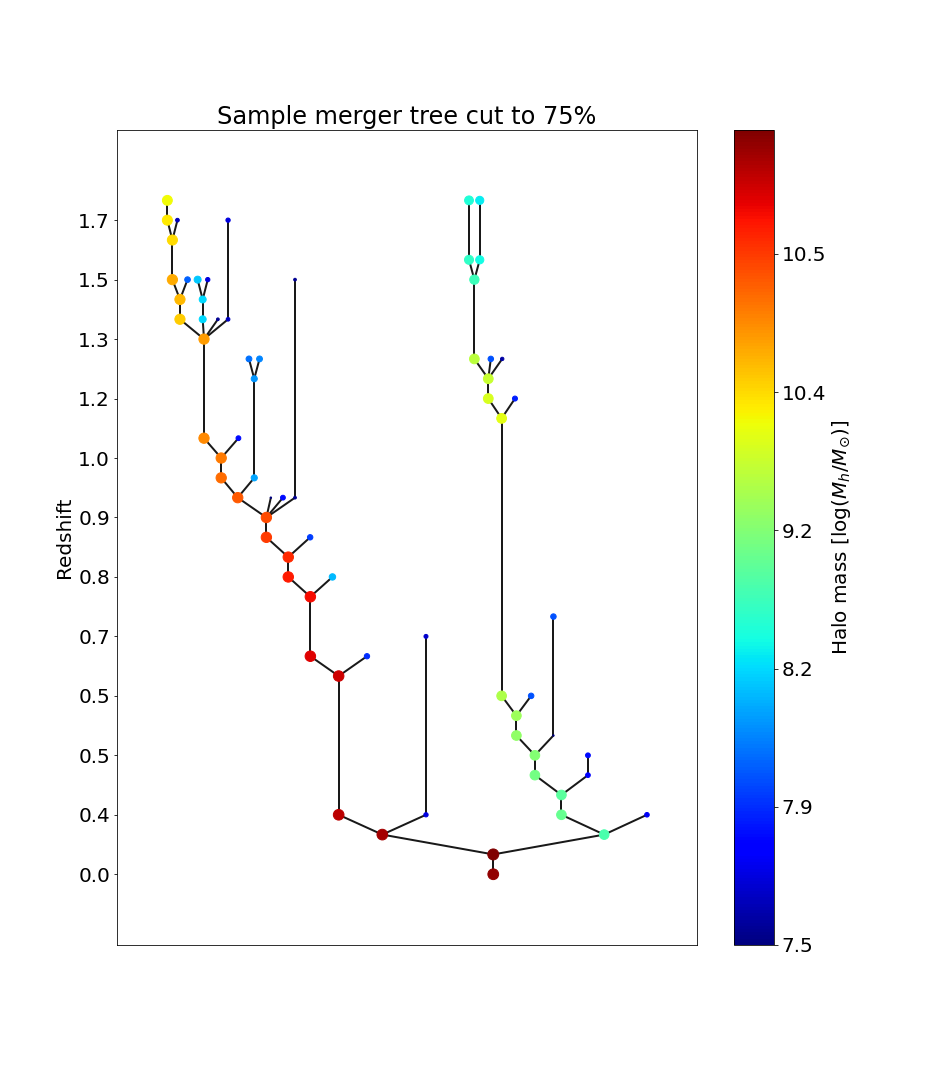}
  \caption{A partial merger tree showing the merger tree from Figure \ref{fig:full_tree} with the nodes with the 75\% highest redshift pruned, which is our method of quantifying the importance of nodes at high redshifts. If our $M_*$ prediction from using this reduced tree is as good as when using the full tree, it indicates that there was no useful information in the 75\% highest redshift nodes. The relation between precision and cutoff percentage can be seen in Figure \ref{fig:trimmed_tree}.}
  \label{fig:partial_tree}
\end{figure}

In order to quantify the dependence on merger history of our galactic stellar mass prediction, we perform a simple study. For all merger trees, we reduce the number of nodes by some fixed percentage, P, for which we retrain and retest the model. We reduce the number of nodes starting at earlier times (high redshift) by finding the scale factor corresponding to the P'th percentile and then excluding all halos at scale factors lower than this. Here we choose the fixed percentage in order to not bias the results towards lower-mass galaxies, since in our dark matter only simulations, their assembly would only start being recorded at lower redshifts, and their merger trees would thus not be pruned to the same extent if one pruned above a fixed redshift. Our chosen pruning method is illustrated in the difference between Figures \ref{fig:full_tree} and \ref{fig:partial_tree}. If our $M_*$ prediction scatter using the P'th - percentile reduced tree is as good as when using the full merger trees, it implies that there was no useful information in the P'th percentile highest redshift nodes that is not also contained in the low redshift nodes.

\begin{figure}[ht]
  \centering
  \includegraphics[width=1.\linewidth]{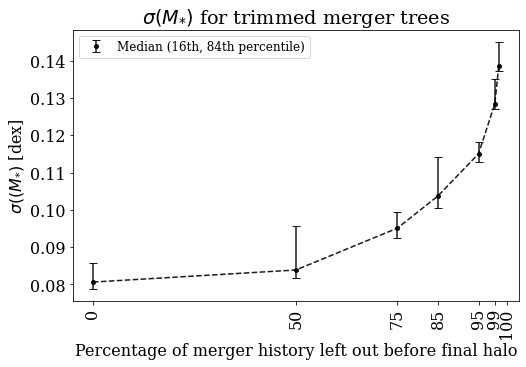}
  \caption{ Median and 16th/84th percentile of the scatter for 25 models at each percentile cut. There is a clear trend towards a loss of accuracy as we remove more nodes, so the stellar mass prediction depends strongly on the assembly history of the given galaxy. The curve is approximately exponential. Note however, that the drop in accuracy when pruning the $50\%$ highest redshift nodes is small.}
  \label{fig:trimmed_tree}
\end{figure}

We choose to train and test 25 times for each percentage. We choose to investigate at 0, 50, 75, 85, 95, 99 and 100\% of the merger history removed\footnote{Removal of 100\% of the merger history corresponds to only using the final halo. Removing 0\% corresponds to using the full merger history.}. The median and 16th/84th percentile of the test scatters are shown in Figure \ref{fig:trimmed_tree}. It is clear that the impact of including more of the merger history is quite significant, with higher relative significance towards halos at lower redshifts.

\vspace{1cm}
\subsection{Feature Ablation}
\label{sec:feature_ablation}

A GNN can act as a predictor of what features are most important in determining the properties of a galaxy. Say that one would like to test whether some connection between black hole mass and three halo parameters exists. Then one can simply train and test a GNN using said three parameters, and compare the results to when the GNN uses all parameters. Thus, in order to investigate which features are the most important for \texttt{Mangrove}'s $M_*$-prediction, we choose this simple approach, where we only include certain sets of features during both training and testing of \texttt{Mangrove}. 

Besides a series of physically motivated sets of quantities (see Table \ref{tab:results_ablation}) from analytical approaches in the literature \citep{Rodriguez_behroozi16_bolshoi_halo}, we also attempt to regress $M_*$ from an \textit{empty tree}, i.e., a tree with no features. The merger tree then contains no information but that encoded in the geometric structure itself. This approach is less precise than the final halo only regression, but it still outperforms abundance matching.

We observe that if one wishes to use only a single parameter, $V_{max}$ has a much lower scatter compared to the otherwise most common choice, the halo mass.

The most salient single group of parameters is the NFW profile parameters.\footnote{Closely followed by $V_{max}$.} The NFW profile information can be encoded either as $M_{200c,500c,2500c}$ or $R_{S,Klypin}$ and $R_{vir}$, which together make the NFW \textbf{concentration parameter} $c_{NFW}$. Combining redshift and NFW profile information rendered the most precise prediction using the fewest features.

Some general interpretability methods rely on the same basic principles of ablation to investigate model behavior, such as SHAP values \citep{Lundberg17_SHAP}. SHAP values are however not reliable for GNNs in their current implementation, but only for simpler models, such as Random Forests \citep{Lundberg18_SHAP}.

\noindent
\begin{table}
\begin{adjustwidth}{-1.3 cm}{}
\begin{tabular}{ l cc cc } 
\toprule
Features used & $\sigma$ & $N_{features}$ \\ 
\midrule
 All & 0.0776 & 37 \\
 Only redshift & 0.1704 & 1 \\
 None/Empty tree & 0.2574 & 0 \\
 Only mass & 0.1436 & 1 \\
 Only NFW\footnote{Navarro-Frenk-White (NFW) profile \citep{NFW_1997}} profile & 0.1082 & 2 \\
 Only $V_{max}$ & 0.1194 & 1 \\
 Redshift and NFW profile & 0.0993 & 3 \\

\bottomrule
\end{tabular}
\end{adjustwidth}
\caption{Results for feature ablation for predicting only $M_*$. Training on a smaller subset of features renders information about the importance of each subset. Interestingly, the empty tree regresses significantly better than abundance matching, demonstrating that there is significant information in just the geometrical structure of the merger tree. We also observe that $V_{max}$ and $C_{NFW}$ are the single parameters that allow \texttt{Mangrove} to make the best predictions.}
\label{tab:results_ablation}
\end{table} 

\section{Discussion and Further Work}
\label{sec:discussionandfurther}

In this section we discuss some of the implications of our work and how these connect to possible future work. Among the topics discussed will be the issues with the constituents of star formation, $M_{cold}$, $SFR$ and $SFR_{100}$, merger history dependence and interpretability, the bias and dispersion in relationships with halo mass, the predicted uncertainties, and how to use \texttt{Mangrove} with other simulations and combining results from \texttt{Mangrove} with observations.

\subsection{$M_{cold}$, $SFR$ and $SFR_{100}$}

Although this paper shows that some highly accurate mappings between dark matter merger trees and baryonic galactic properties exist, there is still significant scatter between the \texttt{Mangrove} and SAM $M_{cold}$, $SFR$ and $SFR_{100}$. It should, however, be noted that the scatters between different SAM runs due to only random seed variation in these quantities are already quite high (see Table \ref{tab:comparison_results}).\\
There were no notable differences in the reconstruction strength between $SFR$ and $SFR_{100}$, which indicates that reconstructing the Star Formation History, as well as the current $SFR$, are similar tasks to the \texttt{Mangrove}.\\
As a way of investigating if \texttt{Mangrove} has learned physically meaningful relationships for these target, we test the interdependence of the target residuals. Here we find that the residuals between the two $SFR$ targets and $M_{cold}$ are strongly correlated (see Figure \ref{fig:residuals} in the appendix), meaning that if \texttt{Mangrove} predicted a too high $M_{cold}$, it would also predict a too high $SFR$, analogous to the Kennicutt-Schmidt relation \citep{Kennicutt88_KSrelation}.\\
The improvement in these three quantities when going from using only the final halo to the full merger history was smaller than expected, since they are thought to be strongly connected to the merger history of the galaxy \citep{White91_hierarchical, SomervilleDave2015, Caplar19_SFMS}. This aspect of galaxy evolution could instead be more strongly connected to their \textit{environments}, which is not included as an input to \texttt{Mangrove}. An obvious possibility for future work is therefore to include an encoding of the environment. Since \citet{Lovell22_hydroERT} showed that summary statistics do not make a significant difference, it would perhaps be wise to include the environmental dependence as an additional graph that extends spatially as in \citet{Makinen22_cosmicgraph}, such that the environmental dependence could be learned. This would naturally lead to including a more explicit subhalo model, with satellites included in the spatial graph. It would then be natural to also predict properties for the satellites. 
\subsection{Merger History Dependence and Interpretability}

For $M_*$, in contrast to $M_{cold}$, $SFR$ and $SFR_{100}$, we find a very strong dependence on the percentage of merger history included, with the importance of including a specific node increasing the closer it is to the $z=0$ (see Figure \ref{fig:trimmed_tree}). This is in contrast to the conclusions of \cite{McGibbon22_ML_nature_vs_nurture}, although their modelling framework does not fully use the formation history.\\
Our analysis of merger history dependence is, however, only preliminary for all parameters but $M_*$. 
The more powerful graph-based framework should facilitate deeper future investigations of the formation history dependence for a wider range of properties. \\
Another interesting avenue for interpreting the model, could be to investigate the model's behavior with symbolic regression \citep{Miles20_symbolicregression}, which would lead to a highly interpretable merger history dependence.
A separate but interesting interpretability approach could be to investigate the merger trees from the point of view of unsupervised learning, as in \citet{Jespersen20_GRB, BerylHovis21_tSNE_redshifts} using either t-SNE or UMAP \citep{VanderMaaten08_tSNE, Mcinnes18_UMAP}. This would be done in the latent space of \texttt{Mangrove}, as the latent space would be more readily comparable.
\subsection{Additional Physical Relationships}

\begin{figure}[ht]
  \centering
  \includegraphics[trim=0.5cm 3cm 0.5cm 3.5cm, clip=true,width=0.95\linewidth]{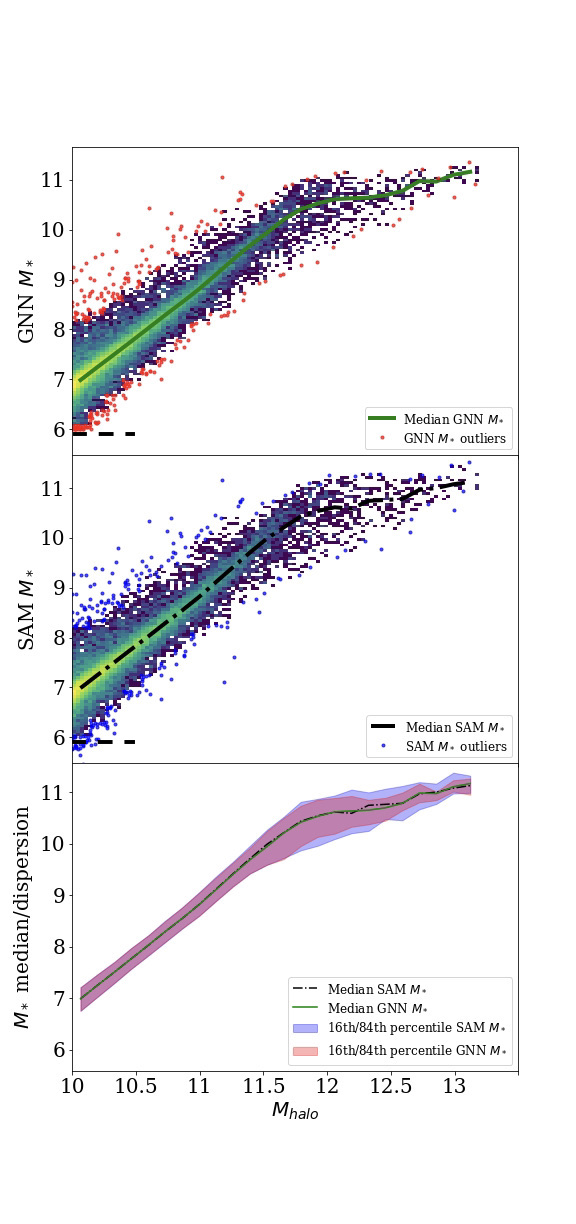}
  \caption{$M_{halo}-M_*$ relation for the SAM (middle), and as predicted by our GNN, \texttt{Mangrove} (upper). For the 99$\%$ of points closest to the median $M_{halo}-M_*$ relation, points are plotted as a histogram with logarithmic colouring, whereas the remaining outliers are plotted as points. The outliers are regressed to a remarkable precision, especially for trees with $M_*$ way above the median $M_{halo}-M_*$ relation. As can be seen comparing the low $M_h, M_*$ region, the Gaussian loss allows \texttt{Mangrove} to ``give up" on some low mass galaxies. This ``limit" is indicated by a dashed black line. Figure \ref{fig:Mhalo-targets} in the appendix shows the same relation for all targets.}
  \label{fig:Mh-mstar}
\end{figure}

As can be seen in Figure \ref{fig:Mh-mstar}, \texttt{Mangrove} reproduces not just the median relationship, but also the dispersion in the $M_{halo}-M_*$ relationship. Figure \ref{fig:Mhalo-targets} in the appendix shows that this is the case for all quantities except the two $SFR$'s. Combining this with the above mentioned correct physical interdependence of the residuals leads to a suspicion that we are closer than ever to being able to emulate the full galaxy property hyperplane. Since \texttt{Mangrove} furthermore performs well at all redshifts, even when interpolating at redshifts where it had not been trained, it is possible to imagine a GNN being able to emulate the full galaxy property hyperplane for all redshifts. Relations like Figure \ref{fig:Mh-mstar} for all other target variables can be found in Appendix \S \ref{asec:more_plots}.\\

Another curious result is the tendency for \texttt{Mangrove} to improve as more output variables were included (with no increase in the amount of model parameters), indicating that an even larger range of baryonic properties than included here could be predicted by one unified model with even greater accuracy, lending further credence to the possibility of emulating the full galaxy property hyperplane.
The authors hypothesize that this is due to \textit{weight smoothing}, which diminishes spurious correlations, since each weight is ``regularized" by the loss from other targets.
\subsection{Uncertainties}

As an additional feature in the literature on emulating baryonic physics, we also have some interesting possibilities due to predicting a set of Gaussian covariances. The diagonal entries, the variance of each parameter, could be used as a filter for selecting predictions that \texttt{Mangrove} is highly confident that it has gotten right. For example, if one filter our results such that only the 50\% highest confidence regressions as judged by \texttt{Mangrove} were included we get significantly better results (see Table \ref{tab:results_50}, and Figure \ref{fig:low_variance} in the appendix). A further investigation into the drivers behind the variance value is beyond the scope of this paper, but highly valuable information could most likely be extracted from these, as demonstrated by \citet{Stiskalek22_halogalxyMLgauss}. Distributions and correlations between the predicted uncertainties and galaxy properties can be found in Appendix \ref{asec:uncertainties}.
\noindent
\begin{table}
\begin{adjustwidth}{-1.2 cm}{}
\begin{threeparttable}
\begin{tabular}{l cc cc cc cc } 
\toprule
Target & $\sigma$ & Bias & $\rho_{Pearson}$ & $R^2$  \\ 
\midrule
 $M_*$ & 0.029 & 0.002 & 0.999 & 0.998 \\
$M_{cold}$ & 0.101 & -0.006 & 0.978 & 0.957 \\
$M_{BH}$ & 0.069 & -0.002 & 0.942 & 0.886 \\
$Z_{gas}$ & 0.085 & -0.008 & 0.987 & 0.975 \\
$SFR$ & 0.217 & 0.017 & 0.974 & 0.948  \\
$SFR_{100}$ & 0.213 & 0.013 & 0.975 & 0.949  \\
\bottomrule
\end{tabular}
\end{threeparttable}
\end{adjustwidth}
\caption{Taking the 50 \% highest confidence regressions as predicted by \texttt{Mangrove}, our results improve greatly, indicating that the predicted variances could be used as a highly efficient filter. A Figure to visualize this improvement can be found in Appendix \S \ref{asec:more_plots}.}
\label{tab:results_50}
\end{table} 

A minor drawback to the generalized Gaussian loss function is that information-sparse areas of the target space can end up being ``ignored" by \texttt{Mangrove}, which instead prefers ascribing a large variance to these points. An example of this can be seen in the halo mass-stellar mass relation in Figure \ref{fig:Mh-mstar}, where low-mass halos with low mass galaxies are effectively ignored in favor of a ``safe" floor value of $M_* \approx 5.9$. Since the issue persists in a mass region where the SAM is already quite uncertain due to the mass resolution, and where the galaxies would not currently be of major importance, this is only of minor concern.

\subsection{Extensions to Other Simulations}

\texttt{Mangrove} can also be applied to the full magneto-hydrodynamical version of IllustrisTNG, as well as any other SAM and hydrodynamical simulation. An analysis of the dependence of baryonic properties on merger history in IllustrisTNG similar to the one presented in this paper would render crucial information as to how and why SAMs and hydrosimulations render different outputs.\\
Since \texttt{Mangrove} works regardless of the number of snapshots included, an interesting avenue to explore could also be different subsampling methods, where specific snapshots are left out according to some scheme, whereafter the \texttt{Mangroves}'s performance on the subsampled merger tree is then measured. This would inform decisions about how many snapshots a simulation team needs to store in order to achieve a satisfactory galaxy property reconstruction, possibly reducing the need for storing an extensive number of snapshots.
\subsection{Connection to Observations}

While extending our framework to other simulations is exciting, we should consider possibilities for combination with observation. Here it should be noted that as determined in \S \ref{sec:assembly_ablation}, only the recent merger history, which is somewhat possible to observe, is required in order to regress the stellar mass significantly more precisely. This kind of recent merger information should be recoverable from spectroscopic missions (e.g. the PFS Galaxy Evolution Survey \citep{PFS14}), since recently merged galaxies can normally be identified from either a strong infrared emission from heated dust, A-star population or $H\alpha$ kinematics \citep{Kennicutt12_SFreview}. Furthermore, we have determined that the most salient information comes from NFW profile information, which is becoming possible to measure \citep{Niikura15_NFWmeasurement}. Halo masses are measurable both from dynamic masses but most confidently from lensing. Therefore, one can imagine a simple population-level version of the model developed in this work being used along with measured stellar masses, recent merger histories and halo features to see if the connections that are emphasized by \texttt{Mangrove} truly are the most important.

\section{Conclusion}
\label{sec:conclusion}

Using the full merger history, we greatly improve upon the current state of the art for emulating galaxy properties with only dark matter properties. We have furthermore shown that interrogating \texttt{Mangrove} renders physical insights into the connection between merger trees and galaxy properties.\\

Considering first only the models use as an emulator, three points are crucial:

\begin{itemize}
  \item \texttt{Mangrove} outperforms all other known ML models when emulating the SC-SAM $M_*$, $M_{cold}$, $Z_{gas}$, $SFR$ , $SFR_{100}$ and $M_{BH}$ by using the full merger history of a given galaxy. Predictions always improve when using the full merger history, and especially $M_*$ and $M_{BH}$ are regressed highly accurately when using the merger history. Since $M_*$ and $M_{BH}$ are regressed so well, they do not just reproduce median relationships, but also the width of the distributions.
  \item When trained, \texttt{Mangrove} is respectively 4 and 9 orders of magnitude faster than the SC-SAM and IllustrisTNG. Including training, this drops to respectively 2 and 7 orders of magnitude. Especially for populating large boxes (side length $>Gpc$), this would drastically reduce run times.
  \item \texttt{Mangrove} works at a range of redshifts and can reliably interpolate between redshifts, even if not trained on galaxies at a given redshift.
\end{itemize}

The physical insights that we have obtained from the model center around two aspects of the connection between merger trees and galactic stellar masses. First, whether the galactic stellar masses are directly related to the properties of the halo within which it resides, or the formation history of the halo, and secondly, which dark matter features are the most valuable. Here we found three especially exciting results.

\begin{itemize}
  \item The earliest half of the merger history of a galaxy can be discarded with only a minor loss of performance when predicting stellar mass. 
  \item Including just 1\% of the merger history closest to the present day leads to significantly improved regression.
  \itemsep1em
  \item \texttt{Mangrove} identifies one especially important set of features, which encode the halos 1-dimensional NFW profile. This can be encoded in two ways, either  by using $M_{200c}$, $M_{500c}$, and $M_{2500c}$ or $R_s$ and $R_{vir}$. The second most important parameter is $V_{max}$.
\end{itemize}

\acknowledgements

The authors would like to thank ChangHoon Hahn, Jiaxuan Li, Michael Toomey, Zach Hemler and David N. Spergel for helpful discussions which greatly improved the paper. The authors would also like to thank the anonymous referee for their helpful and concise comments.
This work used a range of software packages, and among those not cited in the main text are \texttt{NumPy} \citep{numpy}, \texttt{matplotlib} \citep{matplotlib}, \texttt{sklearn} \citep{sklearn}, \texttt{SciPy} \citep{scipy}, \texttt{pandas} \citep{pandas}, \texttt{jupyter} \citep{jupyter} and \texttt{TensorFlow} \citep{tensorflow}.
For external Figures used in Figure \ref{fig:GNN}, simulation Figure credit is due to C. M. Baugh and C. Frenk and halo illustration credit to David Darling.

\bibliographystyle{mnras}
\bibliography{references} 

\appendix

The appendix contains seven sections, concerning the dark matter features used, the construction of the training, validation and test set, an investigation into the predicted uncertainties, further investigation into the relationships of the predictions with halo features and the interdependence of their residuals, as well as a diagram describing the model architecture, a description of the training of the model, and directions for where to obtain our code and data.\\

\section{Dark matter halo features}
\label{asec:params}

In this work we use the following halo features. For a more in-depth explanation, see Appendix B in \citet{Rodriguez_behroozi16_bolshoi_halo}.
\begin{itemize}
  \item Scale: Halo’s scale factor.
  \item Desc Scale: Scale factor of descendent halo, if applicable.
  \item Num prog: Number of progenitor halos—i.e., number of halos at the immediately preceding snapshot that fully merge into this halo.
  \item Mvir: Halo mass, in units of $M_{\odot}$/h.
  \item Rvir: Halo radius, in units of comoving kpc/h.
  \item Rs: NFW scale radius, in units of comoving kpc/h.
  \item Vrms: Halo particle velocity dispersion, in units of physical (i.e., non-comoving) km/s.
  \item mmp?: 1 if the halo is the most-massive progenitor of its descendent halo; 0 if not.
  \item scale of last MM: scale factor of the halo’s last major merger. This is typically defined as a mass ratio greater than 0.3:1.
  \item Vmax: Maximum halo circular velocity
  \item J\_X/J\_Y/J\_Z Halo angular momentum,
  \item Tidal Force: Strongest tidal force from any nearby halo, in dimensionless units
  \item Rs\_Klypin: NFW scale radius in units of comoving kpc/h, determined using Vmax and Mvir.
  \item Mvir\_all: Halo mass, including unbound particles ($M_{\odot}$ /h).
  \item M200m–M2500c: Mass ($M_{\odot}$/h) enclosed within specified overdensities. These include 200m, 200c, 500c, and 2500c, where $\rho$c is critical density and $\rho$m = $\Omega_M \cdot \rho$c is the mean matter density.
  \item Xoff: Offset of halo center 
  \item Voff: Offset of halo center
  \item Spin Bullock: Bullock spin parameter
  \item b to a, c to a: Ratio of second and third largest shape ellipsoid axes (B and C) to the largest shape ellipsoid axis (A)
  \item A[x],A[y],A[z] and A500c[x],A500c[y],A500c[z]: Largest shape ellipsoid axis (comoving kpc/h). 500c indicates that only particles within a specified halo radius are considered.
  \item T/U: ratio of kinetic to potential energies for halo particles.
  \item M\_pe\_Diemer and M\_pe\_Behroozi: Pseudo-evolution corrected halo masses.
  \item Halfmass radius
\end{itemize}

\section{Training, validation and test sets}
\label{asec:train_val_test}
As outlined in \cite{kuhn2013_applied}, it is important that the final model evaluation is made on data that is not used in either the training or for optimizing hyperparameters. Therefore we here split our data in three groups, a training set, a validation set used for evaluating performance during hyperparameter tuning, and a test set for independently evaluating the performance of the final model. The test set is never used during training or hyperparameter optimization.

A 70/10/20 split is used. After optimizing the hyperparameters via the validation set, it is absorbed into the training set for the final training of the models before testing.

\subsection{Training and Testing at Higher Redshifts}
Since all hyper-parameter tuning is done at z = 0, only a training and testing set are constructed for predicting at $z>0$.
For training and testing at $z>0$, it is important to keep in mind that most galaxies at any $z=z_1$ will be a progenitor of a galaxy at $z_2<z_1$. Thus, if one were to naively train a model on baryonic quantities at both $z_1$ and $z_2$ with randomly chosen training and testing sets, there would be significant information leakage from the training to the test set. 

Therefore, we first construct the $z = 0$ dataset according to the above prescription. Next, for a dataset at any $z_n > 0$, for every merger tree, we test if it contains any part of any merger tree in any dataset at a redshift lower than $ z_n $. If it does, we assign it to the set which the descendant galaxy is part of. All merger trees not assigned to either set are then split such that the overall dataset at $z_n$ has an 80/20 split between training and testing.

\section{Uncertainties}
\label{asec:uncertainties}

\begin{figure}[!htp]
  \centering
  \includegraphics[width=.8\linewidth]{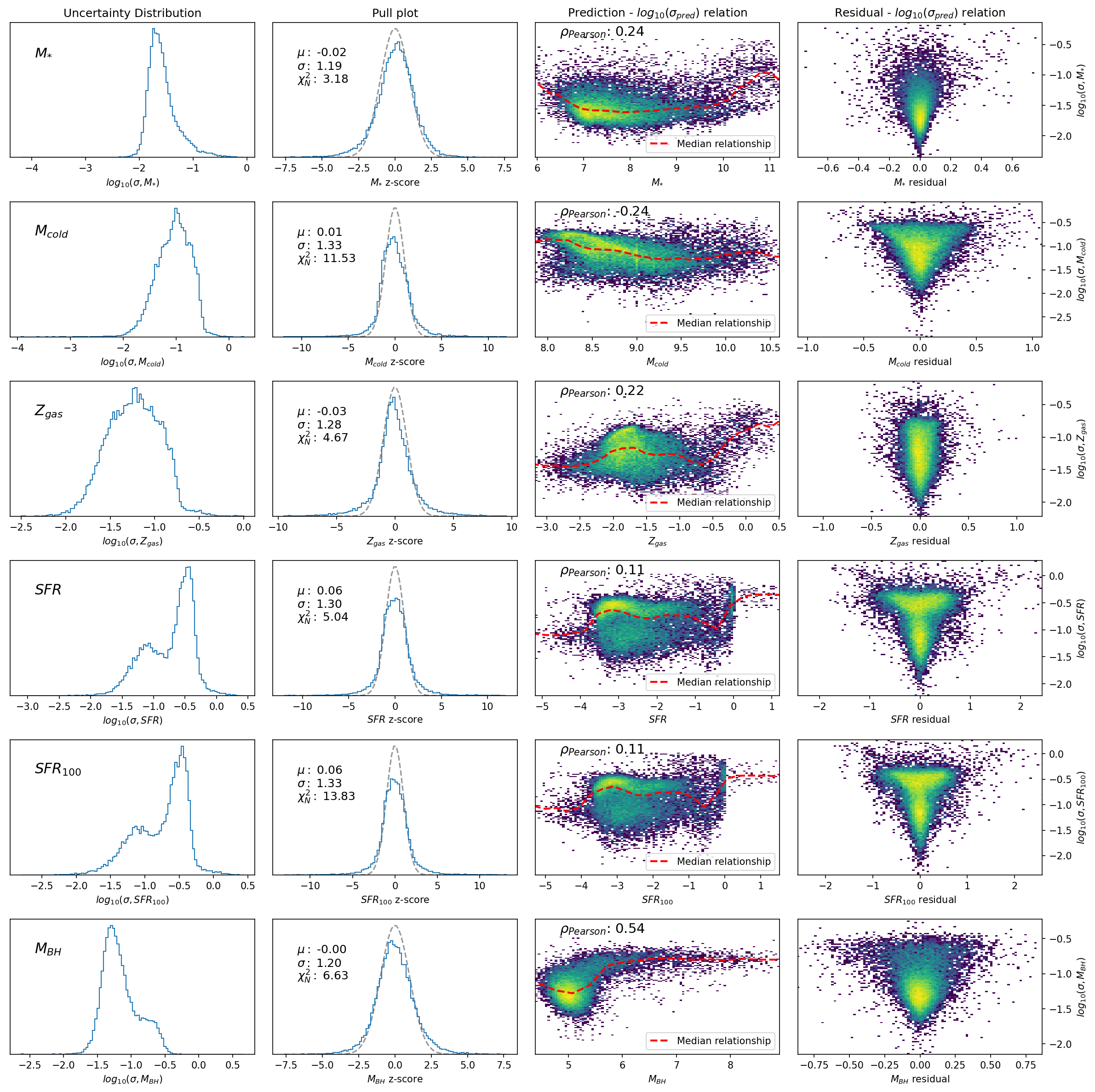}
  \caption{In the first column we show distributions of logarithmic uncertainties. In the second column we show distributions of pulls/z-scores $\frac{\Delta y}{\hat{\sigma}}$, and a unit Gaussian to guide the eye, along with annotations of metrics concerning the pull distributions. The third column shows the relationship between predicted values and uncertainties, along with a red dashed line indicating their median relationship, as well as being annotated with a Pearson correlation coefficient. The fourth and final column shows the relationship between the residuals and the predicted uncertainties, with a clear broadening as one goes toward higher predicted uncertainty.}
  \label{fig:uncertainties}
\end{figure}

The meaning of uncertainties predicted by any neural network are open to interpretation. In a way, they are simply an ensemble spread from the training set, as most networks, including ours, are inherently deterministic. \\
However, since they lack a clear interpretation, we can at least interpret their distributions and accuracies. Since we minimize a Gaussian Likelihood, the pull/z-score distribution ($z = \frac{\Delta y}{\hat{\sigma}}$) should be approximated by a unit Gaussian and the reduced $\chi^2$ ($\chi^2_N=\chi^2/N$) should be close to 1. In Figure \ref{fig:uncertainties}, we show the distributions of logarithmic uncertainties, the distribution of z-scores (pull plot) along with unit Gaussians, as well as relationships between the predictions/residuals and logarithmic uncertainties.Immediately noticeable is the fact that the two $SFR$'s and $M_{BH}$ have strong bimodalities in the uncertainties, and that the uncertainties in general span many orders of magnitude. From the pull plots, we see that all distributions are approximately Gaussian, but with variances $\approx 25\%$ too high, so there is some departure from Gaussianity.
The reduced $\chi^2$ are also too high in general, indicating generally overconfident (too low) uncertainty estimates. It should be noted though, that the $\chi^2$ value \textit{assumes} that the uncertainties are correctly approximated by a Gaussian, which isn't quite true in our case. Therefore, the $\chi^2$ should be used with some caution. We see some correlations between the predictions and the uncertainties, with especially noticeable case being the high uncertainties given to both low and high $M_*$, the high uncertainties given to the highest $Z_{gas}$, $SFR$, $SFR_{100}$ and $M_{BH}$, as well as the the low uncertainties given to higher values of $M_{cold}$.

\newpage
\section{Interpretation plots}
\label{asec:more_plots}

\begin{figure}[!htb]
  \centering
  \includegraphics[width=0.9\linewidth, trim={0cm 0cm 0cm 0cm},clip]{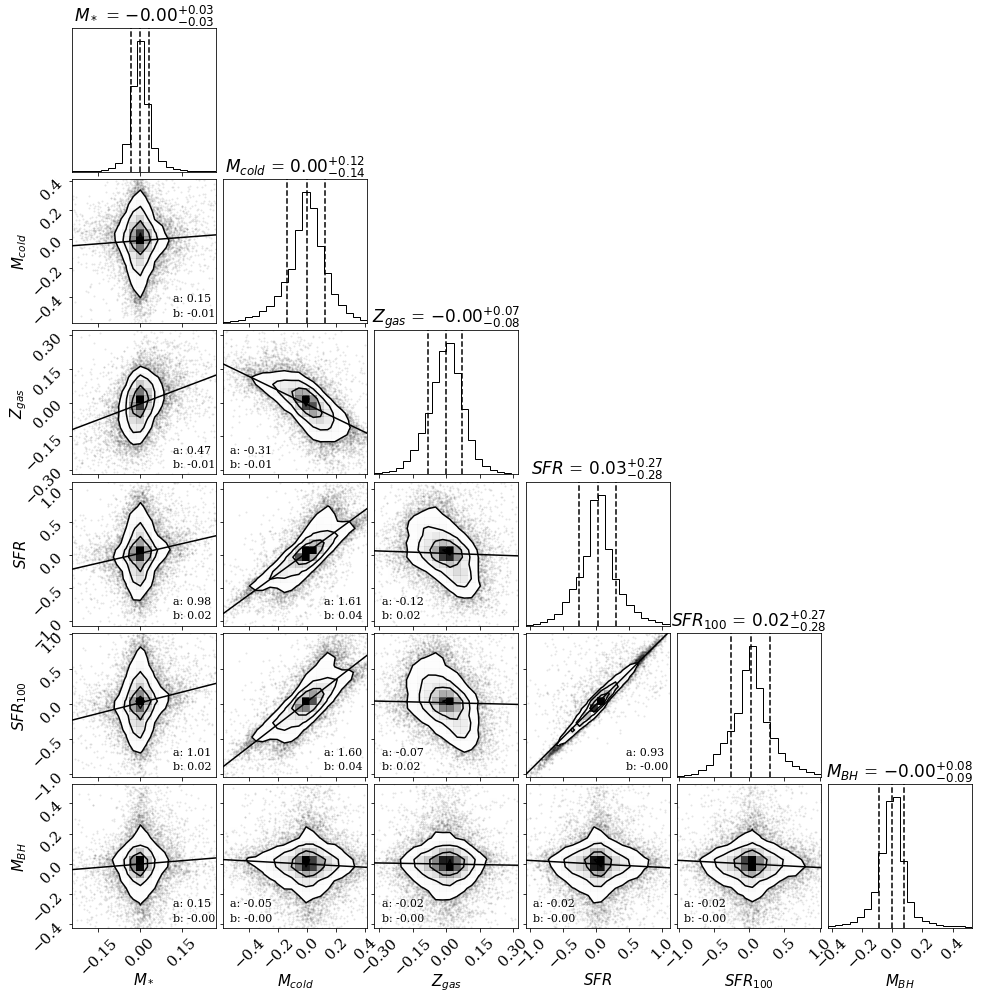}
  \caption{Residual for all targets, along with linear (a*x+b) fits. Each window is annotated with the slope (a) and the intercept (b) of the residual - residual relation in question. The plot is made with the \texttt{corner} package \citep{corner}.}
  \label{fig:residuals}
\end{figure}

There are several checks that we can perform in order to make sure that \texttt{Mangrove} is predicting physically meaningful things, as well as probing the regions where \texttt{Mangrove} struggles. The very tight scatter is highly indicative that most physically relevant relationships will be reproduced, but here we probe these relations further. We follow tests done by \citet{Agarwal:2018, Gabrielpillai21_IllustrisSAM_comparison} for:

\begin{itemize}
  \item Median stellar mass - halo mass deviation dependence on NFW concentration. As shown by G21, this is a property reproduced by both IllustrisTNG and the SC-SAM. This comparison can be found in Figure \ref{fig:concentration}.
  
  \item Halo mass - variable relations (as for stellar mass in Figure \ref{fig:Mh-mstar}) are also generally very useful for identifying the regions where \texttt{Mangrove} fails to reproduce the SAM. This general comparison can be found in Figure \ref{fig:Mhalo-targets}. Here we quickly identify one of the reasons for \texttt{Mangrove}'s poor performance on $SFR$ and $SFR_{100}$, namely that it doesn't successfully capture the two diverging branches of SFR around $M_{halo} \approx 11.7$, regressing only the lower branch accurately. We also observe that $M_*, M_{cold}, Z_{gas}$ and $M_{BH}$ generally follow both the median relation as well as reproducing the scatter. The scatter isn't reproduced for the two $SFR$ targets. This is a problem discussed in \citet{Agarwal:2018}, which our method also improves significantly upon. We also investigate how these quantities evolve within fixed bins of a given halo mass in Figure \ref{fig:Mhalobin-targets}.
  
  \item Residual - residual plots are also very useful for investigating the interdependence between predictions. Here we provide a plot to provide a picture of these interdependences. Figure \ref{fig:residuals}, simply shows residual - residual relations for \texttt{Mangrove} relative to the SAM targets, along with the slope (a) and intercept (b) of a line fitted using least squares (not using the $\sigma$ predicted by \texttt{Mangrove}).\\
  From this plot we clearly observe a strong interdependence between $SFR$ - and $SFR_{100}$ - residuals (as expected), positive correlations between $M_*$ - and $SFR$ / $SFR_{100}$ / $Z_{gas}$ - residuals, positive correlations between $M_{cold}$ and $SFR$ / $SFR_{100}$ - residuals(analogous to a Kennicutt-Schmidt relation) and a negative correlation between $M_{cold}$ - and $Z_{gas}$ - residuals.
\end{itemize}

Besides these tests, we also provide precision Figures in the style of Figures \ref{fig:mstar_performance} and \ref{fig:others_performance} for the performance of the final halo only regression (Figure \ref{fig:final_halo}), as well as for the 50\% lowest variance objects (Figure \ref{fig:low_variance}).

\begin{adjustwidth}{-1.5 cm}{}
\begin{figure}[!htb]
  \centering
  \includegraphics[width=.95\linewidth, trim={4cm 1cm 4cm 1cm}]{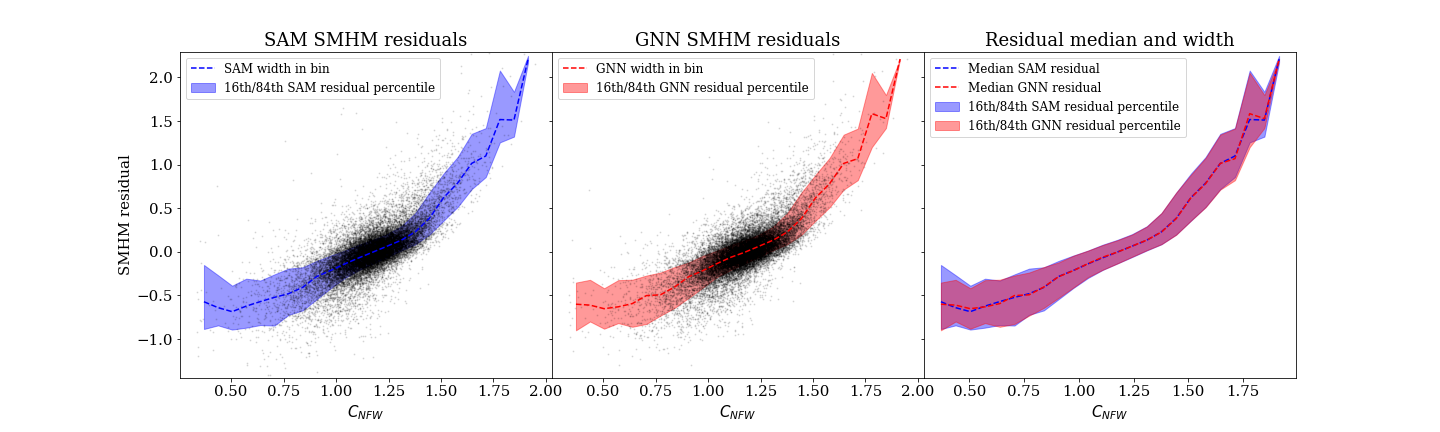}
  \caption{The difference between the $M_*$/$M_{halo}$ and the median value of $M_*$/$M_{halo}$ for halo-galaxy pairs in bins of halo mass bin plotted as a function of the NFW concentration parameter of the halo. The dashed lines show the medians and the shaded areas show the 16 and 84th percentiles. The left panel shows the relationship for the SC SAM, the middle panel shows that for \texttt{Mangrove}, and the rightmost panels shows a comparison between the two. Equal bin widths are chosen for comparison with Figure 12 in G21.}
  \label{fig:concentration}
\end{figure}
\end{adjustwidth}

\begin{adjustwidth}{-3.7 cm}{}
\begin{figure}[!htb]
  \centering
  \includegraphics[width=1.0\linewidth, trim={2cm 7cm 3cm 7cm},clip]{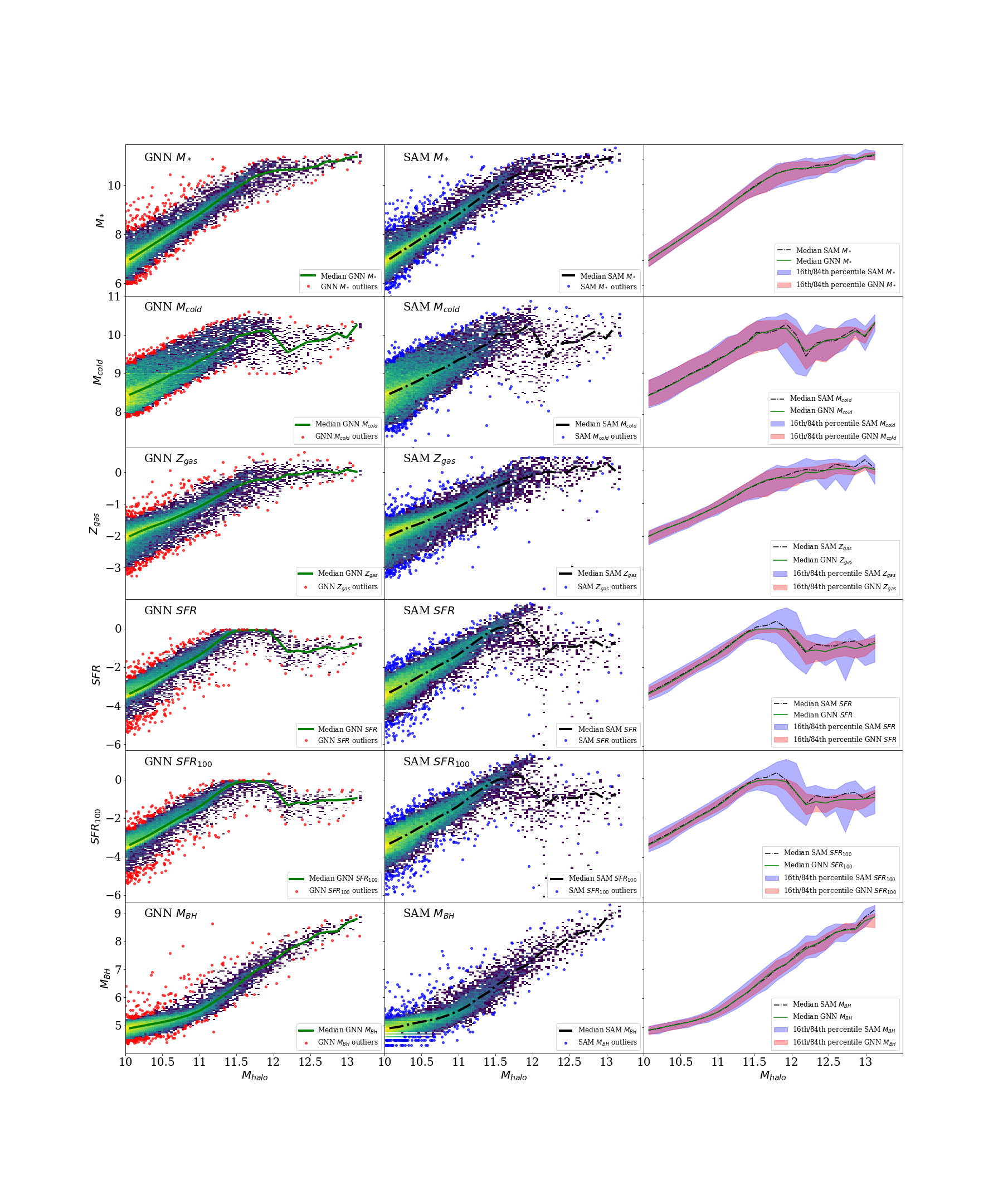}
  \caption{Relation between halo masses and target parameters for all targets for both the SAM and \texttt{Mangrove} predictions, in similar style as Figure \ref{fig:Mh-mstar}, with outliers clearly marked. We furthermore show general trends in the right column, where the dashed and solid lines show the medians and the shaded areas show the 16 and 84th percentiles for the parameter in question for both the SAM and \texttt{Mangrove}. Here we immediately see the source of some of the errors, as for example, the inability of \texttt{Mangrove} to accurately capture the two diverging branches in $SFR$.}
  \label{fig:Mhalo-targets}
\end{figure}
\end{adjustwidth}

\begin{figure}[!htb]
  \centering
  \includegraphics[width=0.85\linewidth, trim={2cm 7cm 5.cm 7cm},clip]{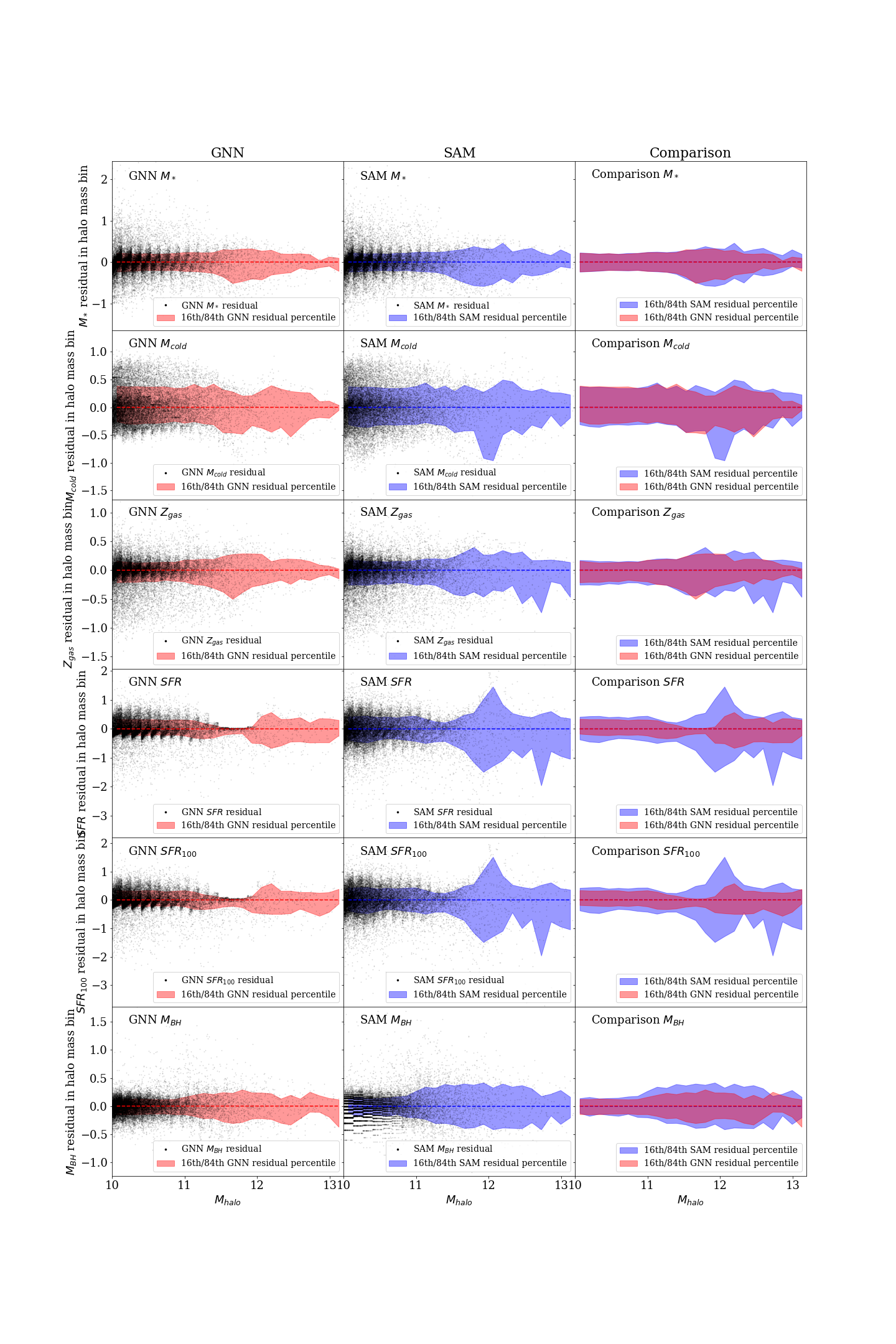}
  \caption{Relation between $M_{halo}$ and target parameter residuals with respect to the median of the target parameters in $M_{halo}$ - bins. 20 bins are used. This is a separate but distinct way of visualizing the main points of Figure \ref{fig:Mhalo-targets}, i.e. that the dispersion is quite accurate for all parameters but $SFR$ and $SFR_{100}$. There is thus hope for producing a full galaxy property hyperplane with \texttt{Mangrove} or a similar GNN - based method. }
  \label{fig:Mhalobin-targets}
\end{figure}

\begin{adjustwidth}{-10 cm}{}
\begin{figure}[!htb]
  \centering
  \includegraphics[width=.9\linewidth, trim={7cm 2cm 5cm 3cm},clip]{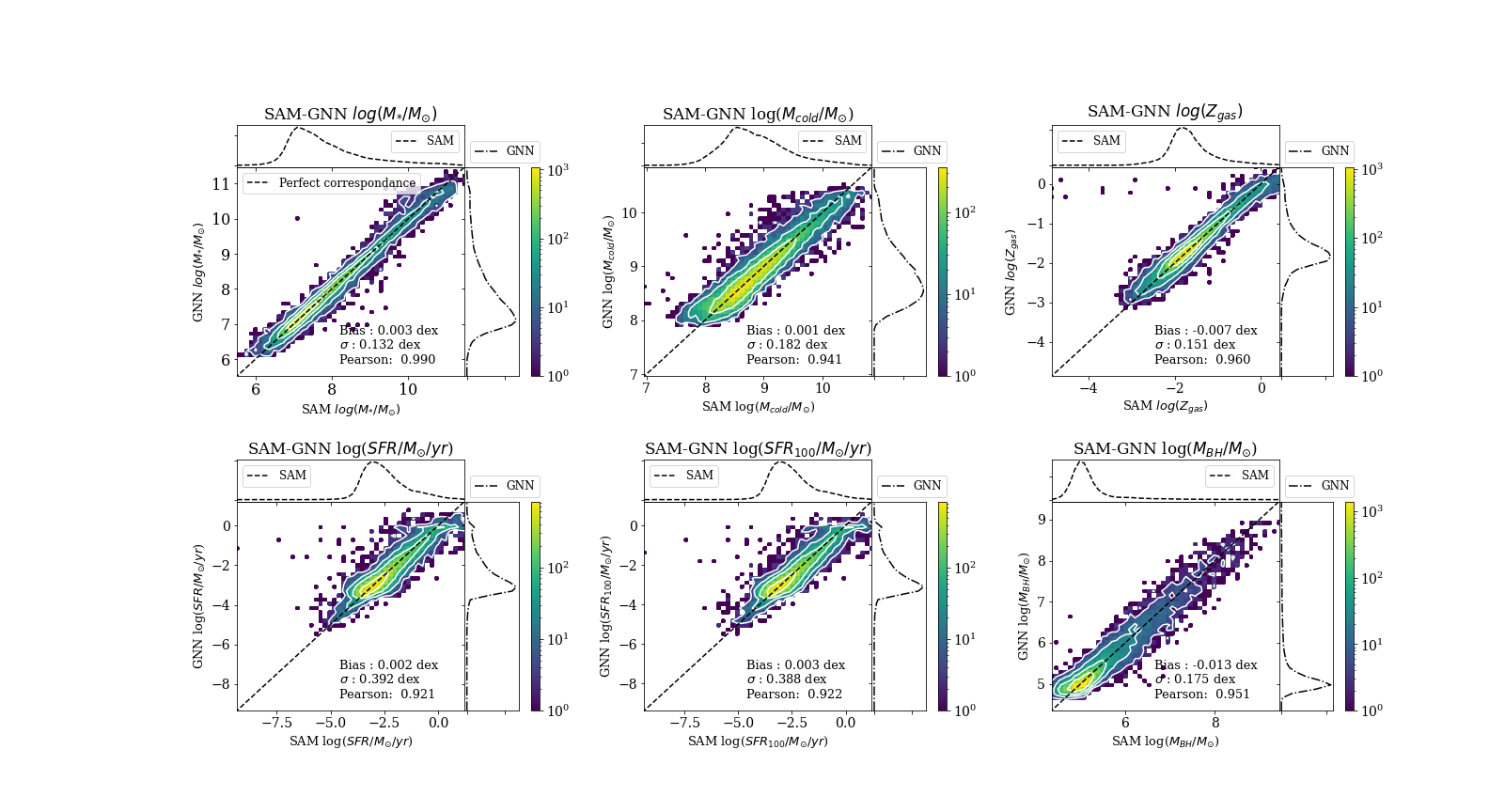}
  \caption{Same as Figures \ref{fig:mstar_performance} and \ref{fig:others_performance}, but for regressing using only information from the final halo. We see a general decline in performance compared to using the full merger tree, although median relations are improved, and the $SFR$ is slightly better when predicting $SFR>0.2$. A single node is still technically a graph, however, no graph structure is used, so the GNN label may be slightly misleading.}
  \label{fig:final_halo}
\end{figure}
\end{adjustwidth}

\begin{adjustwidth}{-10 cm}{}
\begin{figure}[!htb]
  \centering
  \includegraphics[width=.9\linewidth, trim={7cm 2cm 5cm 3cm},clip]{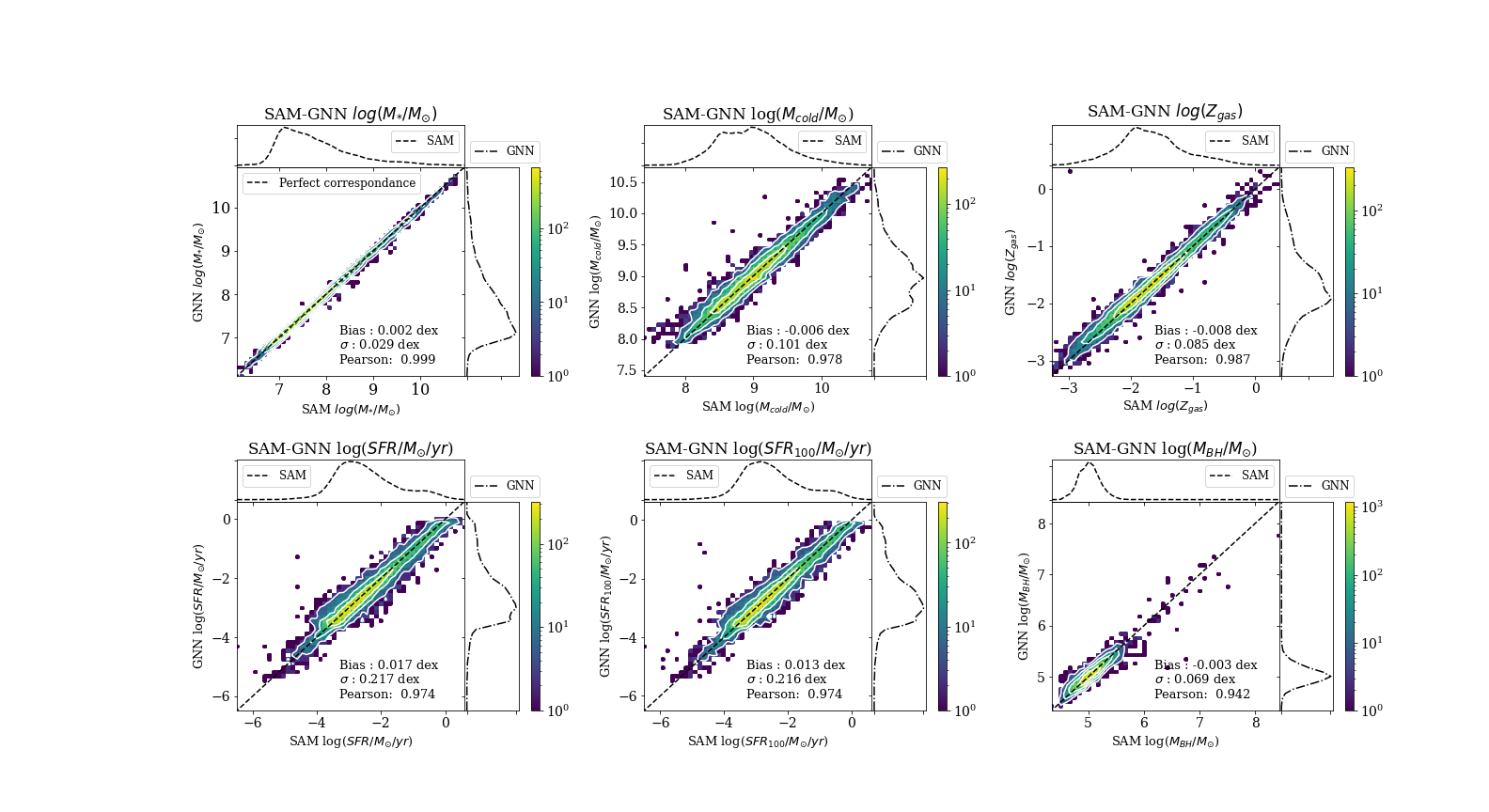}
  \caption{Same as Figures \ref{fig:mstar_performance} and \ref{fig:others_performance}, but using cuts where only the 50 \% lowest variance predictions, as predicted by \texttt{Mangrove}, are included. As can be clearly seen, \texttt{Mangrove} does significantly better, even though it tends to avoid regions that are information - poor, like the massive black hole region ($M_{BH}>6.1$).}
  \label{fig:low_variance}
\end{figure}
\end{adjustwidth}

\begin{adjustwidth}{-10 cm}{}
\begin{figure}[!htb]
  \centering
  \includegraphics[width=.9\linewidth, trim={7cm 2cm 5cm 3cm},clip]{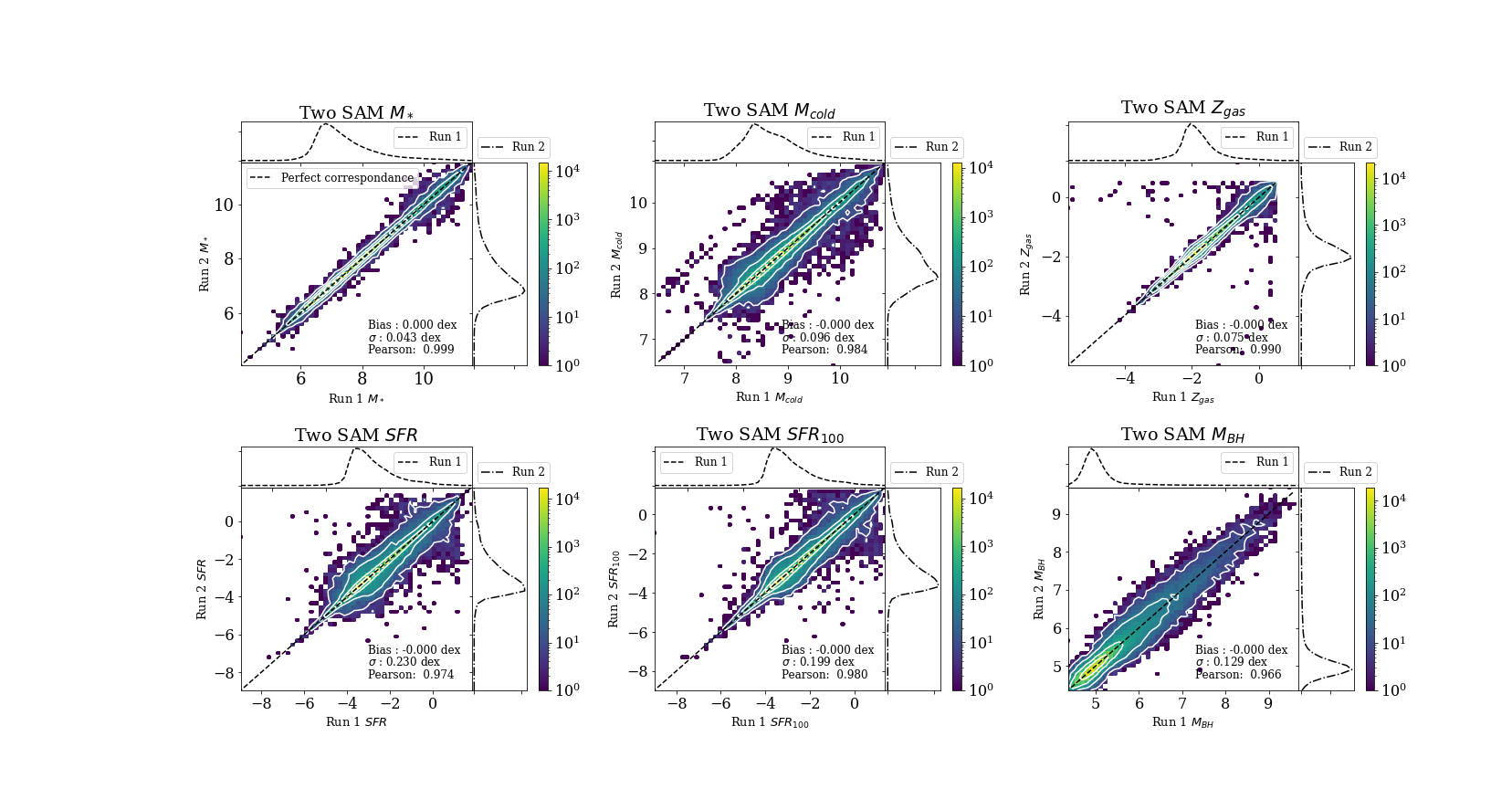}
  \caption{Same as Figures \ref{fig:mstar_performance} and \ref{fig:others_performance}, but where the differences are taken between two different runs of the SC-SAM with nothing changed but the random seeds. The numbers aren't exactly the same as Table \ref{tab:comparison_results}, since those are the mean of the metrics of different SAM runs.}
  \label{fig:SAM_difference}
\end{figure}
\end{adjustwidth}

\clearpage

\section{Model architecture}
\label{sec:architecture}

Here we wish to describe the architecture a bit more in depth for the purpose of reproducibility. The architecture is visualized in Figure \ref{fig:architecture}. The merger tree is passed through a 2-layer Multi-Layer Perceptron (MLP) to encode the node state before any graph convolutional layers. Then the encoded merger tree is passed through 5 GraphSAGE layers, each with a ReLU activation layer between. The encoded merger tree is then summed over with a global sum pooling. Using a global max pooling renders similar performance. Each of the targets then has its own 3-layer MLP decoder ``head". A ``head" means a different branch of the model with all heads taking the same input, allowing each head to predict more independently of the others. If the uncorrelated Gaussian loss is used, no off-diagonal components of the covariance matrix are predicted, and $\hat{\Sigma}$ is diagonal and corresponding to just having the usual Gaussian uncertainties. The layer normalization description can be found in \citet{ba2016_layernorm}. After the sequence of convolutional layers, a differentiable global pooling operator is applied across all nodes in order to standardize the output size. The dimensionality of the latent space (known as the number of hidden states) was 128.

\begin{adjustwidth}{-2.9 cm}{}
\begin{figure}
  \centering
  \includegraphics[width=1.05\linewidth]{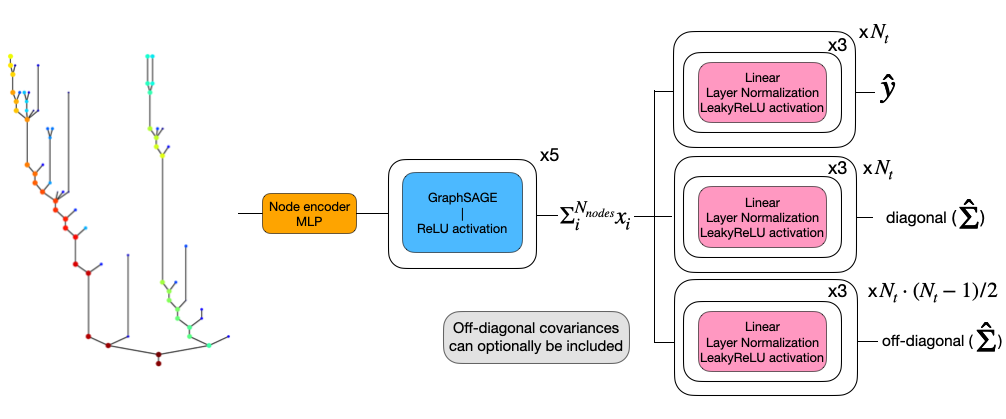}
  \caption{A diagram of \texttt{Mangrove}'s architecture for predicting values and the full-covariance matrix. $N_t$ is the number of targets one wishes to regress. The number of times a given block is repeated is written by the upper right corner of the block. A linear layer is the same as a 1-layer MLP. The flow is from left to right, but inside each box the flow is from top to bottom. Each layer operates with 128 hidden states. The off-diagonal covariances are not included in the work in this paper.}
  \label{fig:architecture}
\end{figure}
\end{adjustwidth}

\section{Training the model}
\label{asec:training}

We train the models using the \verb|Pytorch| OneCycleLR learning rate scheduler \citep{Smith2018_superconvergence_OneCycleLr, Paszke2019_PyTorch}, using a max learning rate of $10^{-2}$ and a batch size of 256 using the Adam optimizer \citep{KingmaBa2017_Adam}. The models were trained for 1000 epochs when optimized for all targets, and 500 for 2 targets or less, as this was determined during hyperparameter\footnote{The hyperparameters of the model and training scheme are defined as parameters not of the model itself, but about the model or training scheme. Examples include the dimensionality of the latent space, the number of layers and the learning rate.} optimization to be above the average number of epochs required for a model to converge. A Gaussian quantile transform \footnote{\href{https://scikit-learn.org/stable/modules/generated/sklearn.preprocessing.QuantileTransformer.html}{sklearn source code}}, which maps each parameter to a Gaussian distribution defined by the quantiles of the parameter in question, was fit on the training set and applied to all input data before training, except for categorical data such as the number of progenitor halos or whether the halo had recently undergone a major merger, which is encoded as a boolean in the data. This makes training more stable at the risk of destroying some information. We also attempted using a standard scaler, which scales data to have zero mean and unity variance. This resulted in slightly higher scatters by about 3-5 $\%$.

We employ a max learning rate of $10^{-2}$, a $15\%$ start percentage and a final division factor of $10^3$. \\

A series of learning rate schedulers (constant, warmup with exponential, cosine annealing and one cycle) were attempted, all with reasonable success. Although the constant learning rate works well, it is suboptimal for long runs, and consistently underperforms by $\approx$ 5\%. Among the others, we generally observe similar performance, although the cosine annealing schedule renders results with higher variance between runs.

\begin{figure}
  \centering
  \includegraphics[width=0.9\linewidth]{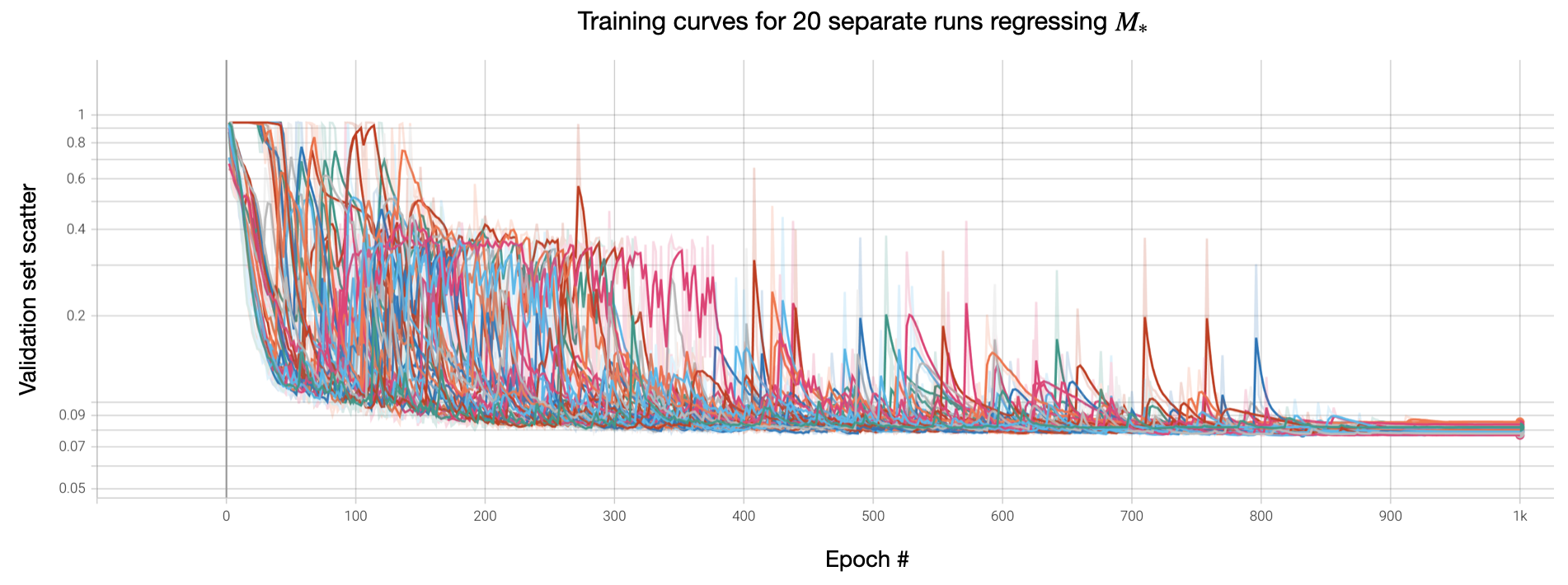}
  \caption{A series of sample training curves for regressing $M_*$ only. Hard lines are smoothed with an exponential kernel with strength 0.6, and shaded are the actual validation scatters. As can be seen, the training can be noisy, and have short, strong spikes, but eventually converges given enough epochs.}
  \label{fig:training}
\end{figure}

\section{Reproducing our results}
\label{asec:code}
The code for reproducing our results can be found GitHub at https://github.com/astrockragh/Mangrove (\href{https://github.com/astrockragh/Mangrove}{\faGithubSquare}). The repository is provided under the MIT license. Data can be obtained from the IllustrisTNG website (https://www.tng-project.org/data/).

\label{lastpage}
\end{document}